\documentclass{openjournal}

\usepackage{xcolor}
\usepackage{textgreek}
\usepackage[utf8]{inputenc}
\usepackage[english]{babel}
\usepackage{rotating}
\usepackage{multirow}
\usepackage{makecell}

\usepackage{natbib}
\defcitealias{xrism-pers}{Velocity Paper}

\usepackage{hyperref}
\hypersetup{
    unicode, 
    colorlinks=true,
    linkcolor=linkcolor,
    citecolor=linkcolor,
    filecolor=linkcolor,
    urlcolor=linkcolor,
}
\usepackage{color,colortbl}
\definecolor{linkcolor}{rgb}{0.0,0.3,0.5}
\usepackage{tensind}
\tensordelimiter{?}
\DeclareGraphicsExtensions{.bmp,.png,.jpg,.pdf}
\usepackage{verbatim}
\usepackage[normalem]{ulem}
\usepackage{orcidlink}
\usepackage{soul}
\usepackage{makecell}
\usepackage{array}
\setcellgapes{4pt}

\graphicspath{ {./figs/} }

\newcommand{\chandra}{{\em Chandra}}
\newcommand{\suzaku}{{\em Suzaku}}
\newcommand{\xrism}{{\em XRISM}}
\newcommand{\hitomi}{{\em Hitomi}}
\newcommand{\resolve}{{\em Resolve}}
\newcommand{\xmm}{{\em XMM-Newton}}
\newcommand{\athena}{{\em NewAthena}}

\usepackage{soul}
\usepackage{amsmath}
\usepackage{xfrac}
\usepackage{capt-of}%

\begin{document}
\title{\xrism{} observations of the Perseus cluster along two arms: Chaotic ICM motions probed by resonant scattering}

\author{Annie Heinrich\orcidlink{0000-0002-7726-4202}$^{1}$}
\email{amheinrich@uchicago.edu}
\author{Irina Zhuravleva\orcidlink{0000-0001-7630-8085}$^{1}$}
\author{Congyao Zhang\orcidlink{0000-0001-5888-7052}$^{2,1}$}
\author{Anna Ogorzalek\orcidlink{0000-0003-4504-2557}$^{3,4,5}$}
\author{Ay\c seg\"ul T\"umer\orcidlink{0000-0002-3132-8776}$^{6,4,5}$}
\author{Fran\c cois Mernier\orcidlink{0000-0002-7031-4772}$^{7,3,4,5}$}
\author{Phillip C. Stancil\orcidlink{0000-0003-4661-6735}$^{8}$}
\author{Elena Bellomi\orcidlink{0000-0001-6411-3686}$^{9}$}
\author{Lior Shefler\orcidlink{0009-0002-3581-5763}$^{8}$}
\author{John ZuHone\orcidlink{0000-0003-3175-2347}$^{9}$}
\author{Yutaka Fujita\orcidlink{0000-0003-0058-9719}$^{10}$}
\author{Julie Hlavacek-Larrondo\orcidlink{0000-0001-7271-7340}$^{11}$}
\author{Yuto Ichinohe\orcidlink{0000-0002-6102-1441}$^{12}$}
\author{Kyoko Matsushita\orcidlink{0000-0003-2907-0902}$^{13}$}
\author{Nhut Truong\orcidlink{0000-0003-4983-0462}$^{6,4,5}$}
\author{Shutaro Ueda\orcidlink{0000-0001-6252-7922}$^{14}$}


\affiliation{$^1$Department of Astronomy \& Astrophysics / University of Chicago}
\affiliation{$^2$Department of Theoretical Physics and Astrophysics / Masaryk University}
\affiliation{$^3$Department of Astronomy / University of Maryland, College Park}
\affiliation{$^4$NASA / Goddard Space Flight Center}
\affiliation{$^5$Center for Research and Exploration in Space Science and Technology, NASA / GSFC}
\affiliation{$^6$Center for Space Sciences and Technology / University of Maryland, Baltimore County}
\affiliation{$^7$Univ Toulouse / CNES, CNRS, IRAP}
\affiliation{$^8$Department of Physics and Astronomy / The University of Georgia}
\affiliation{$^9$Center for Astrophysics / Harvard-Smithsonian}
\affiliation{$^{10}$Department of Physics, Graduate School of Science / Tokyo Metropolitan University}
\affiliation{$^{11}$D\'{e}partement de Physique / Universit\'{e} de Montr\'{e}al}
\affiliation{$^{12}$Nishina Center for Accelerator-Based Science / RIKEN}
\affiliation{$^{13}$Department of Physics, Graduate School of Science / Tokyo University of Science }
\affiliation{$^{14}$Advanced Research Center for Space Science and Technology / Kanazawa University}

\begin{abstract}
    \xrism{} has mapped gas velocities across the core of the Perseus cluster, separating the kinematic effects of mergers and AGN feedback. 
    The physical properties of these motions remain unclear: are they a superposition of bulk flows, predominantly random/turbulent motions, or a mixture of both?
    Without resolving this question, constraints on the nonthermal pressure fraction and heating rate remain uncertain, as both assume predominantly random motions. 
    Unlike emission line broadening, resonant scattering is most sensitive to small-scale, random motions rather than coherent bulk flows.
    {Taking advantage of the extensive \xrism{} coverage of the Perseus cluster, we detect the full effects of resonant scattering on the He$\alpha$ $w$ line for the first time.
    This includes flux suppression in the cluster center, enhancement in the outer regions, and non-Gaussianity in the emission line.}
    We employ radiative transfer simulations to constrain the amplitude of small-scale ICM velocities in the inner $\sim60$ kpc of Perseus, finding them to be consistent with the line broadening measurements within the uncertainties. 
    This indicates the observed velocity dispersion is primarily due to small-scale random motions in the central Perseus regions rather than coherent bulk flows. 
    We further explore potential anisotropy of these motions, showing that they are consistent with isotropic or radial motions rather than tangential ones. 
    Longer \xrism{} observations are required to improve these anisotropy constraints. 
    Finally, we explore azimuthal variations between the two complete radial arms observed by \xrism{}.
\end{abstract}


\maketitle

\section{Introduction}
\label{sec:intro}
Since its launch in 2023, XRISM has measured gas velocities in the intracluster medium (ICM) of more than a dozen galaxy clusters. 
In relaxed clusters, measured velocity dispersions typically fall in the $\sim100$-$300$ km/s range \citep[e.g.,][]{xrism-centaurus,xrism-a2029,xrism-pers,xrism-virgo,majumder2026a,yamada2026}.
Velocities in merging clusters trend higher, often between $\sim200$-$300$ km/s \citep[e.g.,][]{xrism-coma,xrism-a2319,heinrich2025}, with some up to $\sim 400$-$500$ km/s \citep{heinrich2026, omiya2026}. 
Though these velocities are similar between clusters, their physical interpretation depends on their associated scales, which can vary significantly, depending on the cluster core properties and redshift.

As shown in \citet{xrism-pers} \citepalias[hereafter][]{xrism-pers}, the velocity scales probed by XRISM are weighted by the X-ray emissivity of the ICM.
Cool-core clusters have a compact bright center, therefore velocity measurements in these regions are sensitive to smaller scales, e.g., $10-50$ kpc in the Perseus cluster \citep[even $\lesssim5$ kpc in Virgo ]{xrism-virgo}.
Measurements outside of cool-cores and in non-cool-core clusters probe increasingly larger scales with projected radius \citep[see also][]{inogamov2003,zhuravleva2012,ota2026-xrism}.
Thus, resolving ICM kinematics across multiple scales requires spatially mapping gas motions.
Multi-scale velocity measurements are required to distinguish multiple dynamical drivers \citepalias{xrism-pers}, probe velocity cascades \citep[e.g.,][]{xrism-coma,eckert2025,zhang2026}, and estimate the turbulent heating rate in the ICM \citep[e.g.,][]{xrism-virgo,mccall2026}.

Related to the question of scale is the nature of the measured gas motions: are they coherent, bulk flows? Or turbulent, random motions?
The standard picture of a turbulent cascade involves kinetic energy injected on large scales cascading down to progressively smaller scales \citep{kolmogorov1941}.
These small-scale motions are thought to be homogeneous, isotropic, and random.
Calculations of the turbulent heating rate and nonthermal pressure fraction assume this to be true.
As such, the superposition of bulk motions can bias these measurements.
Understanding the properties of the motions probed by observations is therefore critical to many astrophysical applications.

One method that can help to distinguish bulk and random motions is resonant scattering (RS).
While the ICM is generally optically thin, a handful of resonant emission lines can have optical depths of $\tau \gtrsim 1$ \citep{gilfanov1987}. 
For instance, the optical depth at the center of the 6.7 keV Fe He$\alpha$ $w$ is $\sim1$-$3$ in the center of the Perseus cluster \citep{hitomi-rs}.
Therefore, photons at the energies of these lines are expected to scatter out of the line of sight.
 The RS effect makes the $w$ line appear dimmer than expected for an optically-thin plasma in the cluster center and brighter than expected in the outer regions. 
The former effect has been detected in both the atmospheres of giant elliptical galaxies \citep[e.g.,][]{werner2009,deplaa2012,ogorzalek2017} and clusters \citep[][]{hitomi-rs,sarkar2026,tanaka2026}, while the radial trend has never been reported.
Detecting this trend would conclusively demonstrate the presence of RS.

The strength of the RS effect depends on the amplitude of gas motions in the ICM.
A higher velocity (and thus broader emission line) reduces the optical depth/scattering probability, thus regulating the flux suppression/enhancement. 
The photons from the non-resonant, optically thin lines (such as, e.g., the Fe He$\alpha$ forbidden, or $z$ line in the He-like triplet) are not scattered and the line intensities are therefore independent of the velocity from the RS perspective.
Thus, measuring optically thin to thick line ratios via high-resolution X-ray spectroscopy provides an indirect probe of ICM velocities \citep[see][for review]{churazov2010,gu2018,simionescu2019a}.
Beyond the velocity amplitude, \citet{zhuravleva2011} demonstrated that RS probes the scale and anisotropy of the gas motions.
In particular, RS is mostly sensitive to random, small scale motions, such as turbulence.
Combining RS velocity constraints with line-broadening measurements therefore allows us to infer the dominant type of motions.

The Perseus cluster provides the best opportunity to measure ICM kinematics via both line broadening/shifts and RS, {owing to its X-ray brightness and extensive coverage with \xrism{}}. 
Perseus hosts multiple sets of X-ray cavities and large-scale sloshing spirals (Fig. \ref{fig:resid}), suggesting a complex interplay of AGN feedback and mergers \citep[e.g.,][]{fabian2006}. 
This combination of multiple velocity drivers was recently confirmed by \xrism{} by measuring a radial profile of gas velocities in a $\sim260$ kpc arm extending from the cluster center to the northwest \citepalias{xrism-pers}. 
A broader velocity mapping and constraints on the velocity power spectrum in the ICM were subsequently presented in \citet{zhang2026}, while supporting numerical simulations were developed by \citet{bellomi2025}. 
Since the initial \xrism{} analysis, a second arm extending to the south has been fully observed, allowing for the measurement of azimuthal variations in ICM velocities, temperature, and metallicity in addition to radial ones. 
Also, deeper observations of the cluster center are now available.

In this work, we analyze these new \xrism{} observations of Perseus.
Section 2 describes the data reduction, spectral analysis, and radiative transfer simulations we use. 
In Section 3, we present radial profiles of the ICM plasma parameters (temperature, abundances, and velocities), discuss their azimuthal variations, and present a radial profile of the RS effect amplitude. 
In Section 4, we use the observed RS effect and Monte Carlo radiative transfer simulations to measure small-scale velocities in the inner $\sim 60$ kpc region.
Throughout this work, we assume a flat $\Lambda$CDM cosmology with $h = 0.7$ and $\Omega_m = 0.3$. 
At the redshift of Perseus ($z=0.017284$), 1 arcminute corresponds to $\sim22$ kpc.

\section{Methods}
\label{sec:methods}
\subsection{Observations and data reduction}
\label{sec:data}
\begin{figure}
    \centering
    \includegraphics[width=\linewidth]{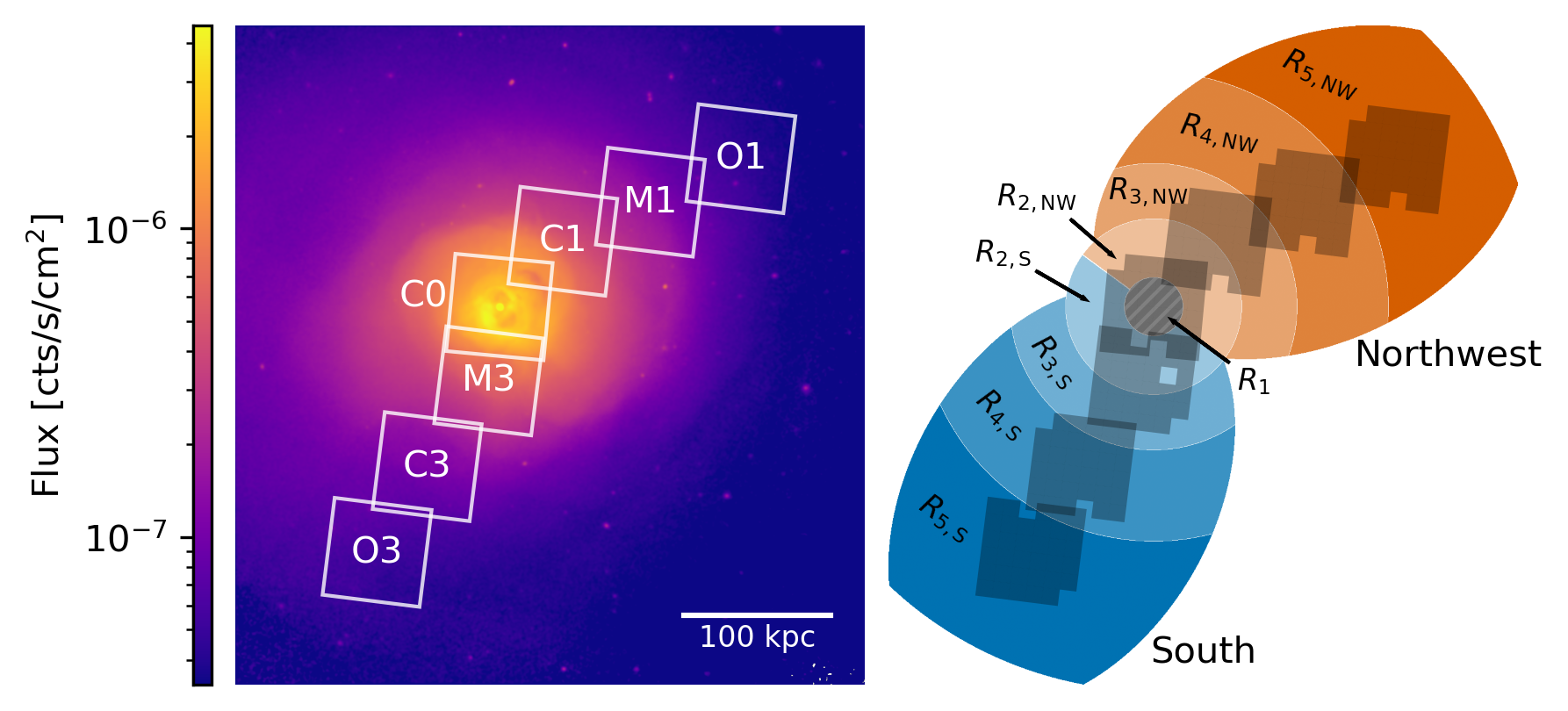}
    \caption{\textbf{Left}: 0.5-8.0 keV exposure-corrected \chandra{} mosaic image of the Perseus cluster, with \xrism{} pointings overlaid as white squares. \textbf{Right:} Sky regions used in this analysis. The central region is shown in hatched grey, the NW arm in orange, and the S arm in blue.}
    \label{fig:binning}
    \vspace{0.3cm}
\end{figure}

Our analysis includes all available \xrism{} observations of two arms covering the core the Perseus cluster as of January 2026.
The observations are split between 7 pointings.
The center of the cluster, C0 (a \resolve{} calibration target) an arm extending to the northwest (C1, M1 and O1; observed during the Performance Verification phase), and an arm extending to the south-southeast (C3, M3, and O3; GO1, PI: Zhuravleva).
C0 includes ObsIDs 000154000, 000155000, 101011010, 101012010, 102007010 and 102008010.
Two other C0 observations (101009010 and 101010010) are available; however, we exclude them from our analysis due to the presence of a significant AGN flare that causes the AGN model to rapidly evolve over the course of the observations (see Appendix \ref{app:agn}).
ObsIDs 000156000, 000157000, and 000158000, 201078010, and 201080010 represent C1, M1, O1, C3, and O3 respectively.
M3 includes both ObsIDs 201079010 and 201079020.
The coordinates, exposure times, and heliocentric velocity of each observation are listed in Table \ref{tab:obsidinfo}.
The \resolve{} fields-of-view (FOVs) of each pointing are shown as white outlines against a broad-band \chandra{} image of the Perseus cluster in Fig. \ref{fig:binning}.

Each ObsID was reprocessed using \texttt{heasoft} version 6.36 and the latest \texttt{CALDB} version (20260315).
Default rise-time screenings were applied to the event files to produce spectra of only high-resolution primary (Hp) events \citep{kilbourne2018}.
Pixel 27 was excluded from each ObsID due to anomalous energy-scale jumps \citep{porter2024}. 
Additionally, pixel 7 was excluded from ObsID 201078010, as this observation was taken while it had degraded resolution\footnote{\url{https://heasarc.gsfc.nasa.gov/FTP/xrism/postlaunch/gainreports/2/201078010\_resolve\_energy\_scale\_report.pdf}}.

Spectra were extracted from several detector regions roughly matching 5 radial bins, extending from the cluster center to the outermost regions observed with our data.
We extracted two spectra from the C0 pointing: one from the central 12 pixels of the \resolve{} array, and another from the outer 26 pixels.
C1 and C3 are also divided into two regions, as the southeastern corner of C1 and the northern half of C3 overlap with the outer region of C0.
Finally, we extracted full-array spectra from M1, O3, M3 and O3, for a total of 27 spectra. 
The pixels used in each region are listed in Table \ref{tab:obsidinfo}.
Large redistribution matrix files were produced for each spectrum using \texttt{rslmkrmf}.
Spectra were then grouped to ensure there is a minimum of 1 count per energy bin.
The spectra from each region are shown in Fig. \ref{fig:spectra}.
In this figure, we have rebinned and stacked the spectra for easier visualization.

We produced auxiliary responses files (ARFs) via the \texttt{xaarfgen} script, which uses ray-tracing simulations to determine the effective area of each detector region given an on-sky source.
Point-source ARFs are generated for the C0 spectra to account for the contribution of the AGN in NGC1275.
For simplicity we ignore the AGN's contribution to all other spectra (C1, C3, M1, M3, O1, and O3), which is $\lesssim2\%$ of the total flux of each of these spectra.
We defined on-sky regions $R_1-R_5$ corresponding to each region from which spectra were extracted, constructed as 5 annuli with radii 0.9', 2.7', 4.4', 7.2', and 11.8' ($\sim20$, 60, 100, 160, and 260 kpc).
All annuli other than the central $R_1$ are split into northwest and southern arms, defined as the overlap of each annulus with two ellipses (with major/minor axes of 14'/9') that cover each arm. 
These regions cover an area on the sky where $\ge 95\%$ of the X-ray photons in each spectrum originate.
ARFs are generated for each spectrum from their corresponding regions using \texttt{xaarfgen} in image mode. 
To account for spatial-spectral mixing (SSM), we additionally generated ARFs for each spectrum from adjacent regions.
Non-adjacent regions contribute $<4\%$ of the light in all spectra, and $\lesssim1\%$ to most spectra.
In every case, we simulated enough ray-traced photons to ensure that $\ge 1000$ photons in each energy bin reach the \resolve{} array footprint. 
For most spectrum/on-sky region combinations, this is accomplished with 1 million photons; however, some require 3 million.

Finally, we extract non-X-ray background (NXB) spectra for each ObsID from the full array using the \xrism{} night-Earth database version 2.
The equivalent screenings were applied to the NXB spectra as the source spectra.
The NXB model comes from \citet{kilbourne2018} and consists of a powerlaw continuum with 12 emission lines. 
We modeled each NXB spectrum independently of the source spectra using a diagonal RMF and no ARF.
We then scaled the model normalization according to the number of pixels from which each spectrum was extracted and fixed this NXB model while fitting the source spectra.

\subsection{Spectral modeling}
\label{sec:modeling}
Each spectrum is modeled in \texttt{xspec} version 12.15.0 using the Cash statistic \citep[C-stat,][]{cash1979}.
The model consists of a combination of NXB, ICM emission from the corresponding on-sky region, and scattered ICM emission from neighboring regions.
AGN emission from NGC1275 ((RA, Dec)=(49.9507$^\circ$, 41.5118$^\circ$)) is also included in the modeling of the C0 spectra.
The appropriate heliocentric correction (see Table \ref{tab:obsidinfo}) is applied to each spectrum via the \texttt{xspec} \texttt{gain} function.

The AGN in NGC1275 is modeled as a single power-law component plus two Gaussians to account for the fluorescent Fe K$\alpha$ lines. 
We consider separate AGN models for each of the 6 central ObsIDs in order to account for possible variation in AGN emission between observations.
As the AGN normalization and ICM continuum normalization are degenerate parameters, we use \chandra{} {measurements of the flux ratio between $R_1$ and $R_2$} to constrain the AGN flux, similar to the procedure used in \citetalias{xrism-pers}.
This procedure is detailed in Appendix \ref{app:agn}.
We fix the fluxes of each AGN model to the fluxes listed in Table \ref{tab:agnpar}, while leaving the AGN spectral index $\Gamma$ free.
{The Fe K$\alpha$ lines have rest-frame energies of 6.404 and 6.391 keV. 
The normalization of the second line is fixed to half the normalization of the first.
The K$\alpha$ line redshifts and widths are tied between observations, while the normalizations are independent.}

We model each sky region as a single absorbed collisionally-ionized equilibrium model (\texttt{tbabs$\cdot$bvapec}), using \texttt{AtomDB} version 3.1.3.
Abundances are relative to the \texttt{lpgs} protosolar abundances table \citep{lodders2009}.
He and C abundances are fixed to solar values, while all other abundances save Ar and Ni are tied to Fe, as these are the only three metals with strong detections in most regions.
As the $w$ line emission is subject to resonant scattering, we remove it from the \texttt{bvapec} model.
In the inner two regions, where the scattering effect is strongest, the $w$ line is replaced with a Voigt line profile in order to account for its modified flux heavier wings.
The normalization {(i.e. the total $w$ line flux)}, Gaussian width $\sigma_w$ and Lorentzian width $\gamma_w$ of the Voigt component are free, while the redshift of the line is tied to the \texttt{bvapec} model.
In the outer three regions, where the scattering effect is weaker, the $w$ line is replaced with a Gaussian component with free normalization. 
In theory, the $w$ line could still be non-Gaussian in these outer regions.
However, we must tie the width of the $w$ line to the width of the \texttt{bvapec} lines in order to measure the velocity dispersion in regions $R_3$-$R_5$ to the greatest precision.
The on-source spectra are fit incrementally: first, all parameters are frozen other than the spectrum normalization and temperature; subsequently redshift, chemical abundances, and velocity dispersion are freed in turn in order to find the baseline ICM models.
{We further ensured that the fit found the global minimum by exploring the parameter space via the \texttt{error} and \texttt{steppar} functions.}
We run two separate fits to the data: one in which all ICM parameters other than redshift and normalization are tied between the NW and S arms to maximize the statistics, and another in which the two arms are independent of one another to investigate azimuthal variations in the cluster.

We quantify the amplitude of the RS effect in two ways.
In the first, we compare the observed $w$ line flux to the flux predicted by an optically thin plasma.
In the second, we measure the line ratios between the brightest optically thin lines in the \xrism{} spectrum and the $w$ line.
Both methods give consistent results.
In this work, we focus on the latter, measuring the Fe He$\alpha_z$ and He$\beta$ to $w$ line ratios.
It is also possible to measure the line ratios of the He$\alpha$ intercombination lines and the Ly$\alpha$ lines; however, the former have known atomic modeling uncertainties \citep{hitomi-atomic,fukushima2026-lines} and the latter are heavily dependent on the ICM temperature gradient (and therefore not particularly useful for probing ICM velocities).
To measure the emission line ratios, we freeze the ICM temperature, metallicity, and velocity dispersion and replace each optically thin line in the \texttt{bvapec} model with a Gaussian line.
The redshifts of each Gaussian are tied to the ICM redshift and the widths to the ICM velocity, while the normalizations are free.

\subsection{Resonant scattering simulations}
We ran Monte Carlo radiative transfer simulations in order to quantify the effect of small-scale gas motions on resonant scattering, following the work of \citet{zhuravleva2010} \citep[see also][]{sazonov2002,churazov2004,churazov2010,hitomi-rs,ogorzalek2017}.
This method has the advantage of allowing for multiple scattering events per photon.
We began with a spherically symmetric model of the Perseus cluster, combining constraints on the ICM density, temperature, and metallicity from \chandra{}, \xmm{}, and \suzaku{} (see Appendix \ref{app:model}).
To account for uncertainties in the deprojected profiles, we randomly sampled the density/temperature/abundance profile to create 100 realizations of the model.
We assume an isotropic turbulent velocity $V_\sigma$ {which is constant} throughout the cluster model.
We additionally tested a model where $V_\sigma$ varies with radius based on the baseline \xrism{} velocity dispersion measurements, finding the difference between the two to be minimal {within the regions considered in this work}.

We simulated a total of 3 billion photons per emission line, which is sufficient to minimize the stochastic scatter of the Monte Carlo simulations.
The initial photon distribution is weighted according to the X-ray emissivity of each radial bin.
Each photon is given a randomly determined propagation direction and an energy drawn from the Doppler-broadened emission line.
We assign an initial weight of $1$ to each photon, representing the cumulative probability of scattering before the photon escapes the cluster.
The optical depth the photon passes through prior to the next scattering event is calculated as $-\ln(1-\epsilon(1-e^{-\tau}))$, where $\epsilon$ is a uniformly distributed random number from 0 to 1.
Each scattering event reduces the photon weight by $1-e^{-\tau}$, where $\tau$ is the optical depth to the edge of the cluster (assumed to be 1 Mpc from the center).
If the photon weight drops below the threshold value of $<10^{-9}$, the photon escapes the cluster.
{Otherwise, the photon is given a new direction and an energy to continue scattering.
The new energy is drawn from the emission line energy distribution at the location of the scattering event.
The new direction is drawn in accordance with the weighted sum of the isotropic and dipole phase functions.
This process is then repeated for subsequent scattering events until the photon weight drops below the threshold.}

The optical depth of the line centroid is calculated as $\tau=\int n_i \sigma ds$, where $ds$ is the photon propagation direction, $n_i$ is the ion number density (calculated from the proton number density, metallicity, and \texttt{AtomDB} ion balance tables), and the scattering cross-section at the center of the line $\sigma=\sqrt{\pi}h c r_e f/\Delta E$, where $f$ is the transition oscillator strength, $c$ is the speed of light, $h$ is the Planck constant and $r_e$ is the classical electron radius.
The Doppler width of the line, $\Delta E$, depends on the line centroid energy $E_0$, the local gas thermal energy $kT$, atomic mass number $A$, proton mass $m_p$, and $V_\sigma$:
\begin{equation}
    \Delta E=\frac{\sqrt{2}E_0}{c}\sqrt{\frac{kT}{Am_p}+V_\sigma^2}
\end{equation}
If $V_\sigma=0$, the optical depths of the $w$, He$\alpha_z$ and He$\beta$ lines are $2.39_{-0.04}^{+0.05}$, $1.01 \pm 0.02 \times10^{-6}$, and $0.386\pm{0.008}$, respectively.
Here, the uncertainties reflect the statistical and systematic uncertainties in the cluster model.

We also ran a set of simulations to explore the potential anisotropy of ICM motions.
These simulations are set up equivalently to the simulations described above; however, we assume a constant 1D velocity dispersion in the ICM, rather than a constant Mach number.
We test three cases: isotropic motions, purely radial motions, and purely tangential motions.
When drawing a new photon energy after each scattering, the width of the line is determined from the velocity component parallel to the photon propagation direction.
When comparing the isotropic, radial, and tangential cases the velocity amplitude is fixed (rather than the kinetic energy).
A detailed description of these simulations is presented in \citet{zhuravleva2011}.

\section{Results}
\label{sec:results}

\begin{figure}
    \centering
    \includegraphics[width=\linewidth]{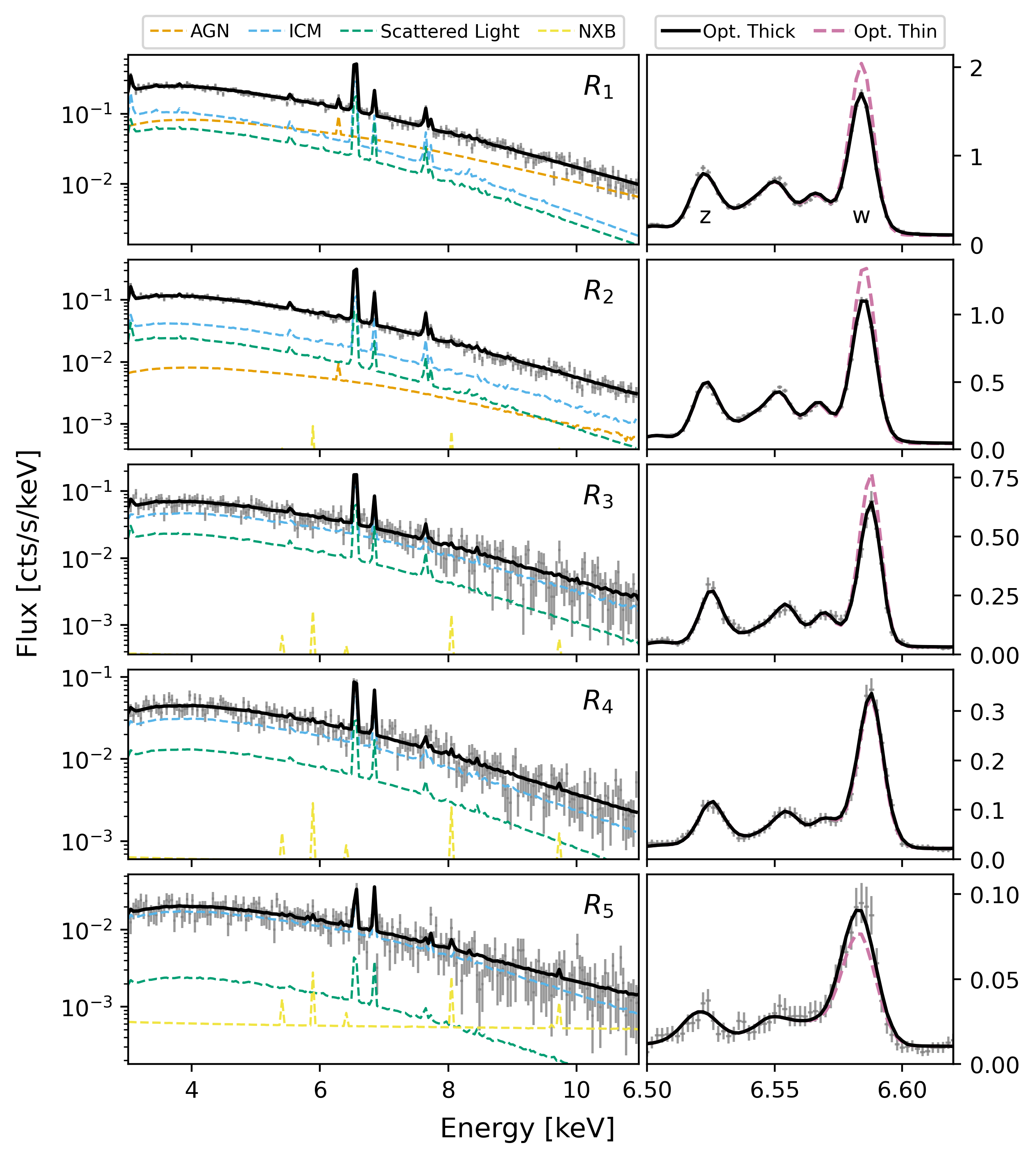}
    \caption{\resolve{} spectra of the Perseus cluster. Spectra from each region (including both arms) are rebinned and stacked for visualization. The left panels show the full 3-11 keV spectra with the best-fit total model in black. The AGN, ICM, scattered light and NXB model components are shown in colored dashed lines. The right panels zoom in on the Fe He$\alpha$ triplet. The black line shows the best-fit model accounting for resonant scattering. The pink dashed line shows the predicted spectrum for an optically-thin $w$ line. In the inner regions ($R_1$-$R_3$) the $w$ line flux is suppressed relative to the optically-thin spectrum, while in the outer regions ($R_4$-$R_5$) the $w$ line is consistent with/brighter than the optically-thin line.}
    \label{fig:spectra}
\end{figure}

\subsection{Radial and azimuthal variations of ICM properties}
\begin{figure}
    \centering
    \includegraphics[width=\linewidth]{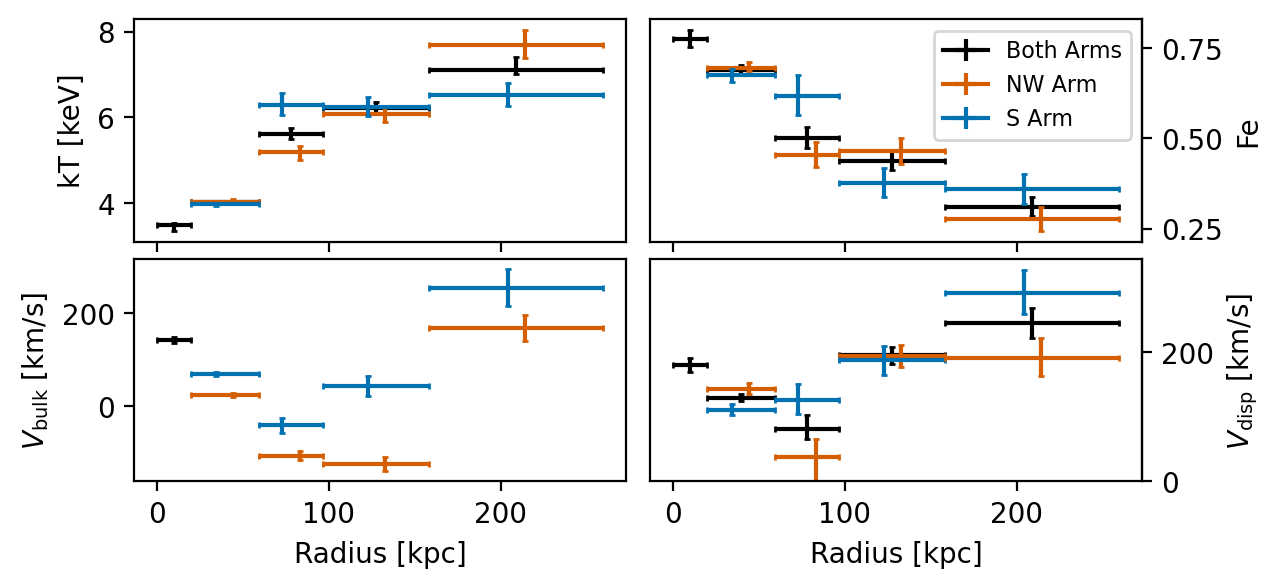}
    \caption{Measured temperature, Fe abundance relative to proto-solar values, bulk velocity and velocity dispersion in Perseus. Orange and blue points denote the NW and S arms, respectively. Black points are the result of the fit when the parameters are tied. Bulk velocities are never tied between arms.}
    \label{fig:icmprofiles}
\end{figure}

Radial profiles of the temperature, Fe abundance, bulk velocity (relative to the redshift of NGC1275, $z=0.017284$), and velocity dispersion are presented in Fig. \ref{fig:icmprofiles}.
The parameters measured when the arms are tied are shown in black, while the independent measurements in the NW/S arms are orange/blue.
The best-fit parameters are also listed in Tables \ref{tab:icmpartied} and \ref{tab:icmparfree}.
Additionally, we map the bulk velocity/velocity dispersion in Fig. \ref{fig:velmap}.
Leaving the two arms independent of one another improve the fit by a $\Delta$C-stat of $\sim70$.
However, this improvement is not significant enough to justify the increased model complexity according to the Bayesian information criterion ($\mathrm{BIC}=k\ln(n)+\text{C-stat}$, where $k$ is the number of free parameters and $n$ is the number of spectral bins).
According to this metric, the tied-arms model is favored by a $\Delta$BIC of $\sim 168$, where $\Delta\text{BIC}\gtrsim10$ is considered significant.
Therefore, we consider the tied-arms model as our ``default'' model.

When the arms are tied, we find a monotonically increasing/decreasing temperature/metallicity.
The cental Fe abundance is {$0.76\pm0.03$}, though we note there is a systematic uncertainty in our AGN model which could cause the central abundance to vary between {$\sim 0.69-0.91$} (discussed in Appendix \ref{app:agn}) 
This measurement is consistent with the metallicities measured by \citet{xrism-pers-metals}, which treats this systematic uncertainty in the AGN model in more detail.
Other than the central abundance, our measured profiles are consistent with those found by \citetalias{xrism-pers} (which adopted an AGN model with a lower flux).
{We also note that the metallicity we measure in $R_3$-$R_5$ is slightly lower than metallicity in the same regions measured by \chandra{} and \xmm{} (Fig. \ref{fig:perseusmod}).
This may be due to a combination of RS enhancing the $w$ line flux in these regions (Section \ref{sec:rsprofile}), incomplete azimuthal coverage with \resolve{}, or the fact that the \chandra{} and \xmm{} profiles measure deprojected metallicities.}

We find a distinctly ``V'' shaped velocity dispersion profile, consistent with previous results in Perseus and expected for gas motions driven by a combination of gas sloshing and AGN feedback \citep{xrism-pers,bellomi2025}.
This profile is inconsistent with a constant velocity dispersion at the $>8\sigma$ level.
To demonstrate this, we fit the tied-arms profile with a broken linear model, finding the break between the decreasing and increasing regimes (which can naively be considered the feedback- and sloshing-dominated regions) at a radius of $\sim 65 \pm 20$ kpc from the cluster center.

We measure similar bulk velocity profiles in the S and NW arms, which are both redshifted relative to the BCG near the cluster center, blueshifted at intermediate radii, and redshifted at large radii.
This pattern is consistent with bulk motions driven by gas sloshing, as concluded in \citetalias{xrism-pers} and \citet{zhang2026}.
The southern ICM is always redshifted with respect to the northwest arm, suggesting the presence of large-scale, sloshing-induced coherent/rotational motions.
This result is consistent with measurements of the bulk velocity structure in Perseus via Cu-K$\alpha$-calibrated \xmm{} spectra \citep{sanders2020}.

When the two arms are fit independently, we find several notable differences between them.
Rather than continually increasing, the temperature profile of the southern arm plateaus at $\sim 6.4$ keV beyond 60 kpc.
This is consistent with \chandra{} and \xmm{} temperature maps of Perseus \citep{churazov2003}, which show a mostly isothermal region between 60 and 260 kpc south of the cluster center, likely caused by gas sloshing.
The highest velocity dispersion is found in the outermost annulus $R_5$.
This is expected given the large scales contributing to the measurement in this region \citepalias[$\sim200$ kpc,][]{xrism-pers}.
However, we also see that motions in the southern arm ($R_\mathrm{5,S}$) are stronger than in the equivalent region of the NW arm: $\sim290$ km/s as opposed to $\sim190$ km/s.
This may be due to the S arm's proximity to the bright eastern region of the cluster, which has enhanced gas velocities, possibly due to large-scale gas sloshing \citep{zhang2026}.

Both the metallicity and velocity dispersion are $\sim2\sigma$ higher in the S than the NW of $R_3$.
This may indicate that AGN feedback is uplifting metals and driving velocities at higher radii in this region than the $\sim 60$ kpc region found by \citetalias{xrism-pers}.
This may be because the sphere of AGN feedback influence is constrained in the NW by the prominent cold front at radius of $\sim60$ kpc (Fig. \ref{fig:resid}).
No cold front is found at the same radius in the S arm, thus AGN feedback may be allowed to extend its influence into $R_3$.
In addition, there is a well-known ``bay'' structure that lies $\sim$between $R_\mathrm{3,S}$ and $R_\mathrm{4,S}$ (see the lower left hand corner of Fig. \ref{fig:resid})
The bay could be the rim of an ancient AGN-inflated bubble; {however, radio emission has not been detected within the bay \citep{groeneveld2026}}.
If so, this would provide an obvious explanation for the observed abundance/velocity enhancements in $R_\mathrm{3,S}$.

It has also been suggested that the bay is a Kelvin-Helmholtz instability in a sloshing cold front \citep[KHi,][]{fabian2006,walker2017}.
If that is the case, sloshing itself could be responsible for the uplifted metals and generation of turbulence in this region.
We also note that the ICM is blueshifted in $R_\mathrm{3,S}$ and redshifted in $R_\mathrm{4,S}$, with respect to the BCG.
This velocity shear is expected if the bay is a sloshing KHi.
If the bay is an AGN cavity this would indicate the wake has ``detached'' from the buoyantly rising bubble \citep{zhang2022a}.
Our results cannot rule out either scenario; further observations and theoretical work is needed to determine the nature of this structure.

\begin{table}[]
    \centering
    \makegapedcells
    \begin{tabular}{c|cccccccc}
    Region & kT [keV] & Ar & Fe & Ni & $V_\mathrm{bulk}$ [km/s] & $V_\mathrm{disp}$ [km/s] & $\sigma_w$ [km/s] & $\gamma_w$ [km/s]\\ 
\hline\\ 
$R_\mathrm{1}$ & $3.38_{-0.06}^{+0.15}$ & $0.85_{-0.08}^{+0.13}$ & $0.76_{-0.03}^{+0.02}$ & $0.7_{-0.2}^{+0.1}$ & $+142_{-6}^{+7}$ & $175 \pm 11$ & $150 \pm 20$ & $110_{-50}^{+40}$\\ 
$R_\mathrm{2}$ & $4.03_{-0.03}^{+0.02}$ & $0.62_{-0.06}^{+0.05}$ & $0.69 \pm 0.01$ & $0.71 \pm 0.07$ & \makecell{$+24_{-4}^{+5}$ \\ $+70 \pm 5$} & $130 \pm 6$ & $158_{-9}^{+8}$ & $50 \pm 20$\\ 
$R_\mathrm{3}$ & $5.6 \pm 0.1$ & $0.8_{-0.3}^{+0.2}$ & $0.5 \pm 0.03$ & $0.4_{-0.1}^{+0.2}$ & \makecell{$-109_{-10}^{+11}$ \\ $-50_{-10}^{+20}$} & $80_{-10}^{+20}$ & --- & ---\\ 
$R_\mathrm{4}$ & $6.2_{-0.1}^{+0.2}$ & $0.4_{-0.2}^{+0.3}$ & $0.44_{-0.03}^{+0.02}$ & $0.5 \pm 0.2$ & \makecell{$-120 \pm 20$ \\ $+40 \pm 20$} & $200_{-20}^{+10}$ & --- & ---\\ 
$R_\mathrm{5}$ & $7.1_{-0.1}^{+0.3}$ & $0.6 \pm 0.3$ & $0.31 \pm 0.03$ & $<0.4$ & \makecell{$+170_{-30}^{+40}$ \\ $+250_{-40}^{+30}$} & $240_{-20}^{+30}$ & --- & ---
    \end{tabular}
    \caption{Best-fit ICM parameters when the NW and S arms are tied. Bulk velocities are relative to the redshift of NGC1275, 0.017284. As the redshift is not tied between arms, we list the bulk velocity for both arms (NW first, S second). $\sigma_w$ and $\gamma_w$ are the Gaussian and Lorentzian widths of the Voigt profile used to model the $w$ line in $R_1$ and $R_2$.}
    \label{tab:icmpartied}
\end{table}

\begin{table}[]
    \centering
    \makegapedcells
    \begin{tabular}{c|cccccccc}
    Region & kT [keV] & Ar & Fe & Ni & $V_\mathrm{bulk}$ [km/s] & $V_\mathrm{disp}$ [km/s] & $\sigma_w$ [km/s] & $\gamma_w$ [km/s]\\ 
\hline\\ 
$R_\mathrm{1}$ & $3.48_{-0.14}^{+0.06}$ & $0.91_{-0.09}^{+0.08}$ & $0.78_{-0.03}^{+0.02}$ & $0.7_{-0.2}^{+0.1}$ & $+142_{-6}^{+7}$ & $181 \pm 11$ & $160_{-20}^{+10}$ & $100 \pm 40$\\ 
$R_\mathrm{2,NW}$ & $4.02_{-0.04}^{+0.07}$ & $0.6 \pm 0.08$ & $0.695_{-0.009}^{+0.018}$ & $0.65 \pm 0.09$ & $+24_{-4}^{+5}$ & $144 \pm 8$ & $160 \pm 10$ & $60_{-30}^{+20}$\\ 
$R_\mathrm{2,S}$ & $3.97 \pm 0.04$ & $0.61_{-0.08}^{+0.09}$ & $0.67_{-0.01}^{+0.02}$ & $0.77_{-0.1}^{+0.09}$ & $+69_{-4}^{+5}$ & $111_{-8}^{+9}$ & $160_{-20}^{+10}$ & $50 \pm 30$\\ 
$R_\mathrm{3,NW}$ & $5.2_{-0.2}^{+0.1}$ & $0.9_{-0.3}^{+0.2}$ & $0.45_{-0.03}^{+0.04}$ & $0.4 \pm 0.2$ & $-107_{-9}^{+10}$ & $<70$ & --- & ---\\ 
$R_\mathrm{3,S}$ & $6.3_{-0.2}^{+0.3}$ & $0.6_{-0.5}^{+0.4}$ & $0.62_{-0.06}^{+0.05}$ & $0.5_{-0.3}^{+0.2}$ & $-40 \pm 20$ & $130_{-30}^{+20}$ & --- & ---\\ 
$R_\mathrm{4,NW}$ & $6.1 \pm 0.2$ & $0.7_{-0.3}^{+0.4}$ & $0.46_{-0.03}^{+0.04}$ & $0.8 \pm 0.2$ & $-130_{-10}^{+20}$ & $190_{-10}^{+20}$ & --- & ---\\ 
$R_\mathrm{4,S}$ & $6.2_{-0.2}^{+0.3}$ & $<0.46$ & $0.38 \pm 0.04$ & $<0.4$ & $+40_{-20}^{+30}$ & $190 \pm 20$ & --- & ---\\ 
$R_\mathrm{5,NW}$ & $7.7_{-0.3}^{+0.4}$ & $<0.5$ & $0.28_{-0.04}^{+0.03}$ & $<0.5$ & $+170 \pm 30$ & $190 \pm 30$ & --- & ---\\ 
$R_\mathrm{5,S}$ & $6.5_{-0.2}^{+0.3}$ & $1.1_{-0.4}^{+0.6}$ & $0.36 \pm 0.04$ & $<0.39$ & $+260 \pm 40$ & $290_{-30}^{+40}$ & --- & ---
    \end{tabular}
    \caption{Best-fit ICM parameters when the NW and S arms are independent.}
    \label{tab:icmparfree}
\end{table}

\subsection{Radial profile of resonant scattering}
\label{sec:rsprofile}

Resonant scattering is expected to reduce the flux of the He$\alpha$ $w$ line in the dense, central regions of the ICM.
The photons from the cluster center will then be redistributed out of the core, increasing the flux of the $w$ line in these regions.
While the central flux suppression has been detected in Persues \citep{hitomi-rs} and PKS0745 \citep{tanaka2026}, the reciprocal flux enhancement in the outskirts has yet to be detected.
The right panels of Fig. \ref{fig:spectra} shows the best-fit model of the Fe He$\alpha$ triplet in black, with the $w$ line of an optically thin plasma plotted as a pink dashed curve.
It is readily apparent that in the inner three regions, the $w$ line flux is suppressed with respect to the optically thin case due to resonant scattering.
This effect appears most strongly in the dense central region, as expected.
The best-fit model is consistent with the optically thin case in $R_4$, while in $R_5$ the observed $w$ line is brighter than the optically thin prediction.
This result is the first demonstration of RS redistributing flux from the cluster center to the outer regions, following theoretical expectations of the behavior of resonant lines in the ICM.

The ratios of the observed (optically thick) $w$ line flux to the optically thin prediction are $\sim 0.81 \pm 0.05$, $0.90\pm0.03$, $0.81\pm0.10$, $1.2\pm0.1$, and $1.2\pm0.2$ in regions $R_1$-$R_5$.
The $w$ line flux is clearly suppressed in the $R_1$ and $R_2$, while it is broadly consistent with the optically thin expectations in all others.
{There is some variations between the S and NW arms: most notably, the optically thick/thin ratio is $1.15 \pm 0.16$ in $R_\mathrm{5,NW}$ and $1.37 \pm0.17$ in $R_\mathrm{5,S}$, where the enhancement in the $w$ line flux due to RS is detected at $\sim2\sigma$ significance.}
We note that Fig. \ref{fig:spectra} shows the $w$ line consistent with optically thin predictions due to SSM.

{While we correct for SSM between adjacent regions, we consider the possibility that the apparent light scattering from the cluster center to the outer regions is due to the PSF rather than RS.
We run an additional fit while correcting for SSM between each region, its neighbors, and its neighbors-of-neighbors.
We find the contribution of light to $R_3$ from $R_1$ due to SSM is $\lesssim1\%$, thus the SSM effect to $R_4$ and $R_5$ is negligible.
We can therefore confidently conclude that the observed scattering of $w$ line flux from the center to the outer regions is due to RS, rather than SSM.}

\begin{figure}
    \centering
    \includegraphics[width=\linewidth]{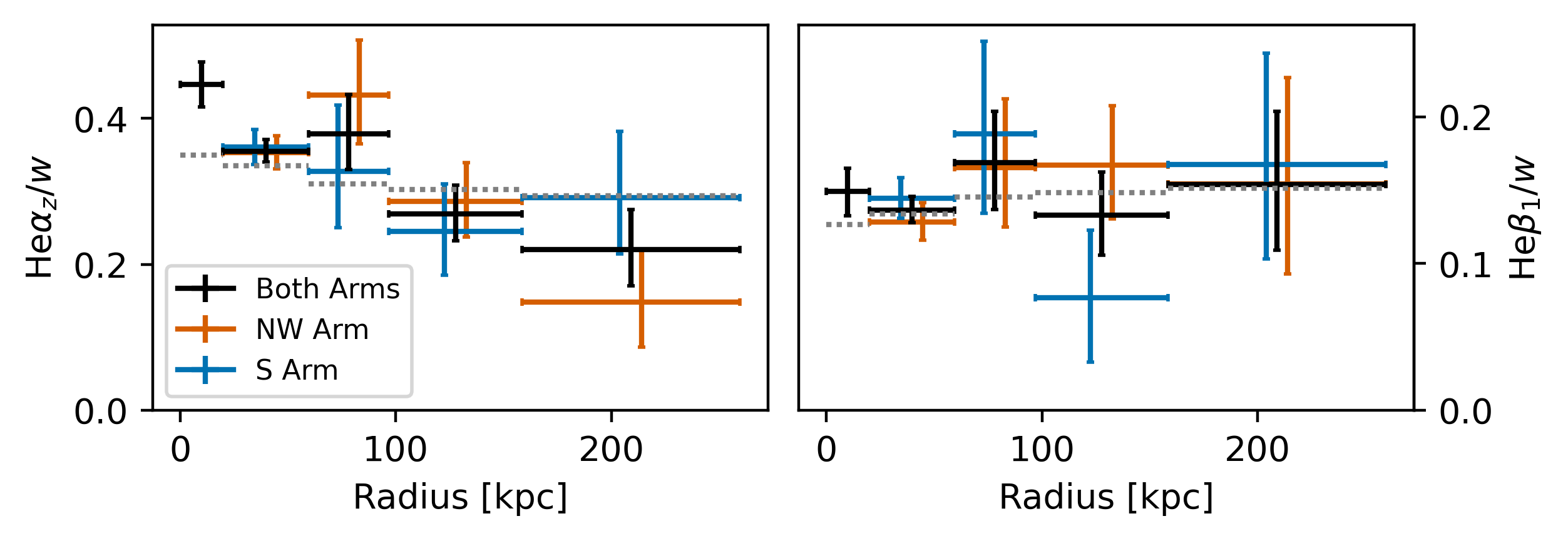}
    \caption{Radial profile and azimuthal variations of the $z/w$ and $\beta/w$ line ratios. Notations are equivalent to Fig. \ref{fig:icmprofiles}. We show the predicted line ratio for an optically thin plasma (using \textsc{AtomDB}) of the same temperature as a gray dotted line.}
    \label{fig:lineratios}
\end{figure}

To further demonstrate this, we measure the line ratios of the optically thin He$\alpha_z$ and He$\beta_1$ lines to the resonant $w$ line (hereafter $z/w$ and $\beta/w$).
The radial profiles of these line ratios are presented in Fig. \ref{fig:lineratios}.
We additionally plot as dotted grey lines the expected line ratios for an optically thin plasma, calculated via \textsc{AtomDB}.
This further shows the RS effect detection in the central $\sim 20$ kpc region, where the $z/w$ ($\beta/w$) ratio is $\gtrsim 3\sigma$ ($1.4\sigma$) greater than the optically thin prediction.
All other line ratios are consistent with an optically thin plasma within $\lesssim 1.4\sigma$.

\section{Discussion}
\label{sec:discussion}

\subsection{Resonant scattering simulations: isotropic gas motions}
\label{sec:rsiso}

\begin{figure}
    \centering
    \includegraphics[width=\linewidth]{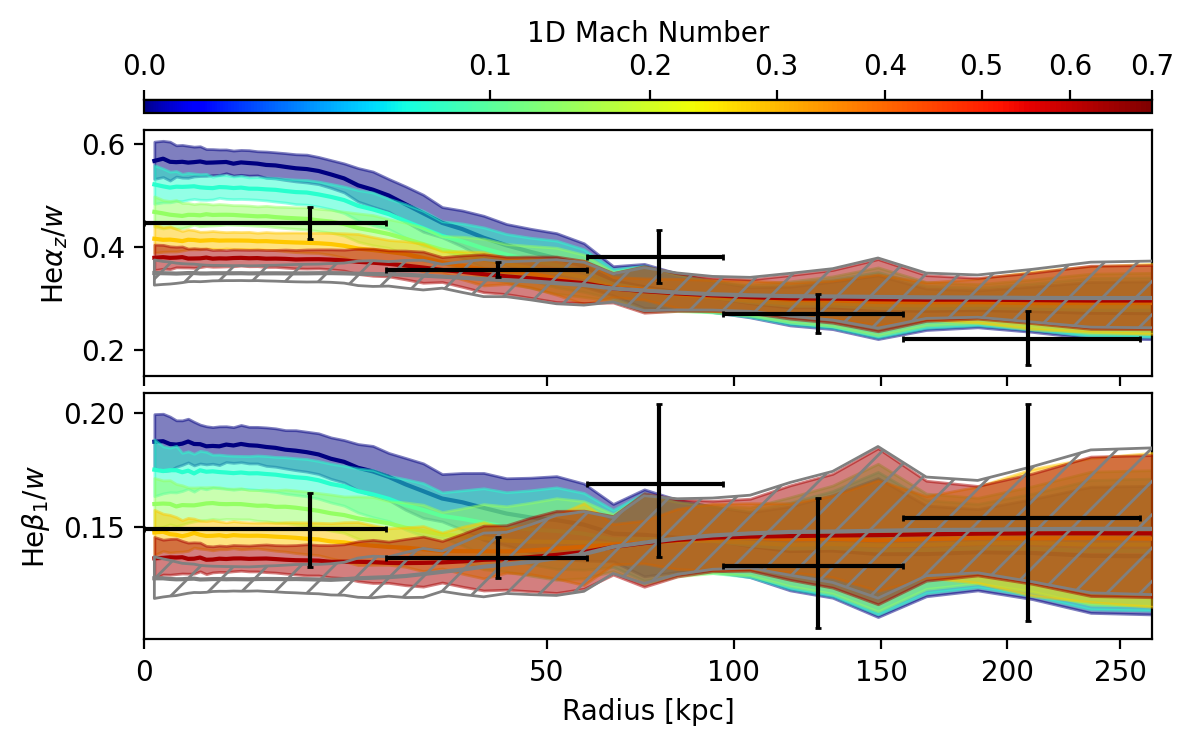}
    \caption{Simulated line ratio profiles for He$\alpha_z$/$w$ and He$\beta$/$w$. Colored curves show the line ratio assuming different values of the turbulent Mach number. Purple, blue, green, yellow, and red curves show the line ratios for $\mathcal{M}_\mathrm{1D}\approx0$, $0.07$, $0.14$, $0.28$, and $0.64$, respectively. Shaded regions denote 1$\sigma$ systematic uncertainties, which are associated with uncertainties in the deprojected cluster model. The hatched grey region shows the expected line ratio in the absence of resonant scattering (the optically thin case). Black crosses show the observed line ratios and $1\sigma$ statistical uncertainties.}
    \label{fig:rssims}
    \vspace{0.2cm}
\end{figure}

In Fig. \ref{fig:rssims}, we compare the measured line ratios to those predicted by our Monte Carlo resonant scattering simulations.
Line ratios predicted for different Mach numbers are shown in different colors, with shaded regions denoting $1\sigma$ uncertainties.
These uncertainties come primarily from the systematic differences between \xmm{} and \chandra{} (see Fig. \ref{fig:perseusmod}).
We also show in a gray hatched region the expected line ratio without scattering.
As expected, the RS effect is weaker in simulations with a higher Mach number, yielding line ratios closer to the optically thin case.
The RS effect is strongest (and thus most useful for velocity constraints) in the dense cluster center.

To derive constraints on the gas velocity from the observed line ratios, we first construct an image of the cluster for each emission line from the radial profiles shown in Fig. \ref{fig:rssims}.
Then, we find the total flux of each emission line in the regions described in Section \ref{sec:methods}, and calculate the line ratios for each region.
This provides an X-ray surface-brightness weighted average line ratio in each region, which are then compared to the observed line ratios.
We are able to use these line ratios to derive velocity constraints in $R_1$ and $R_2$; however, the outer regions are consistent with all simulated velocities because the RS effect is negligible in these regions.

\begin{figure}
    \centering
    \includegraphics[width=\linewidth]{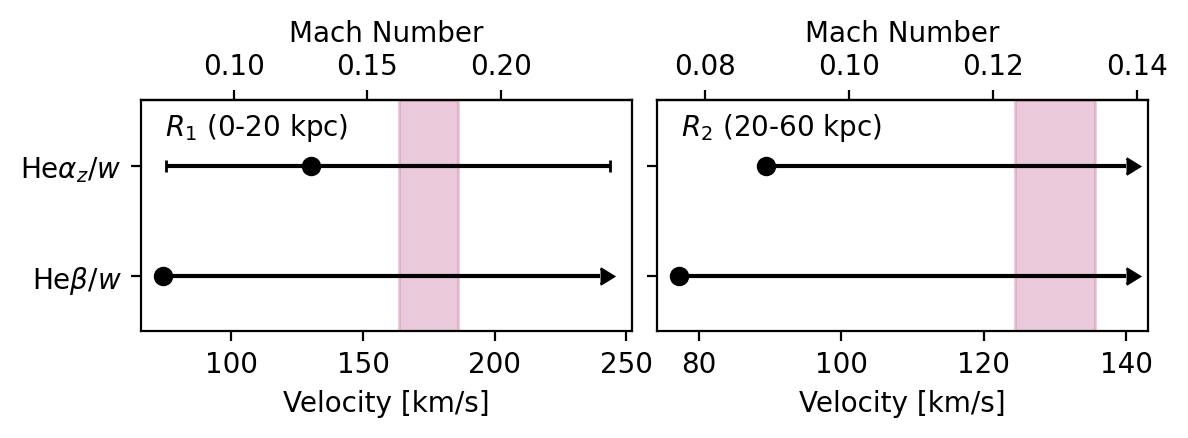}
    \caption{Resonance scattering-derived velocity constraints from the He$\alpha_z$/$w$ and He$\beta$/$w$ ratios (black). Uncertainties and lower limits are $1\sigma$. Pink shaded regions are the velocity measurements derived from Doppler broadening.}
    \vspace{0.12cm}
    \label{fig:vellims2}
\end{figure}

These velocity constraints are plotted in Fig. \ref{fig:vellims2} in comparison with the velocity dispersion measured via Doppler broadening in pink.
In the central region $R_1$, the velocity constrained by the $z/w$ ratio is $\sim 130_{-55}^{+115}$ km/s.
This corresponds to a central optical depth of $\sim 1.2 \pm 0.5$.
In the next region $R_2$, we find a $1\sigma$ lower limit on the velocity of $\gtrsim 90$ km/s.
The velocity constraints derived from the $\beta/w$ line ratio are consistent with these measurements.
Each of these constraints are consistent with the velocity dispersion measured via Doppler broadening.
Using these velocity constraints, we calculate the kinetic pressure fraction in the ICM $\left( 1+3\gamma^{-1}\mathcal{M}_\mathrm{3D}^{-2}\right)^{-1}$, given the adiabatic index $\gamma=5/3$ and the 3D Mach number $\mathcal{M}_\mathrm{3D}=\sqrt{3}V_\sigma/c_s$.
In $R_1$, we find a kinetic pressure fraction of $\sim6_{-4}^{+13}\%$, while in $R_2$ we place a lower limit of $\gtrsim2.5\%$.
Our results are consistent with a similar measurement by \hitomi{} \citep[$\sim150_{-50}^{+80}$,][though we note the regions analyzed with \hitomi{} are slightly different]{hitomi-rs}.

Line broadening is sensitive to both small-scale random motions and coherent motions along the line-of-sight while RS is predominantly sensitive to small-scale, random motions \citep{zhuravleva2011}.
In both $R_1$ and $R_2$, the random motions probed by RS are consistent with the velocity dispersions {measured from line broadening}.
This indicates that the measured velocity dispersions are primarily due to chaotic, small scale motions, rather than bulk flows.

We stress that comparing these two velocity measurement methods is nontrivial, as they may also be sensitive to gas motions on different scales.
The \resolve{}-measured spectrum is weighted by the X-ray emissivity, which is $\propto n_e^2$.
As the ICM is stratified, the spectrum is dominated by emission from a region near the cluster center with a size $\ell_\mathrm{eff}$ \citepalias[the effective length scale {as defined in}][]{xrism-pers} that sets the largest scale of gas motions Doppler broadening is sensitive to.
However, the RS effect is additionally weighted by the ion density along the LoS, which is $\propto n_e$.
How this additional weighting affects the scales probed by RS is not immediately clear and is beyond the scope of this work.

\subsection{Resonant scattering simulations: anisotropic gas motions}
\label{sec:rsaniso}

Thus far we have assumed that small-scale, random gas motions in Perseus are isotropic.
However, it is possible that the gas motions are preferentially radial or tangential.
The former case may be driven by expanding AGN bubbles or shocks, while the latter could be due to internal gravity waves {generated by gas sloshing or buoyantly rising bubbles \citep{zuhone2022,zhang2018}}.
Previous work has shown that the RS effect is sensitive to the directional anisotropy of gas motions in addition to their amplitude and scale \citep{zhuravleva2011}.
We therefore run a grid of radiative transfer simulations with purely radial and tangential motions in addition to the default isotropic case.

\begin{figure}
    \centering
    \includegraphics[width=\linewidth]{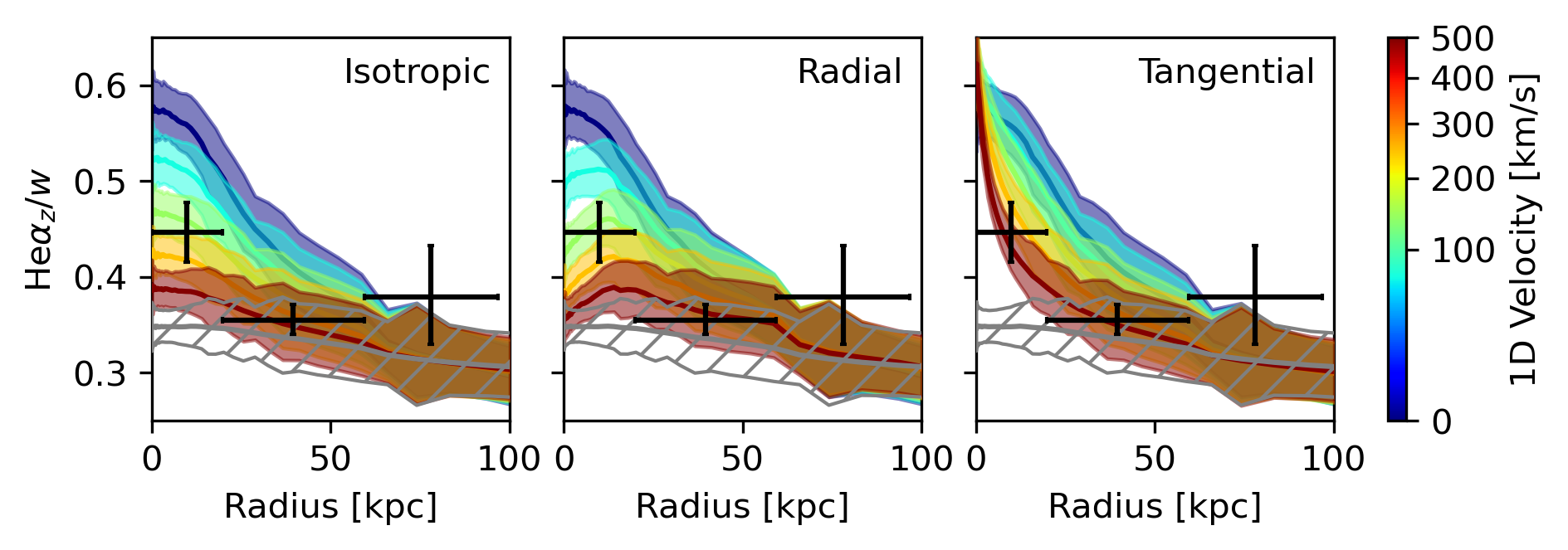}
    \caption{Simulated line ratio profiles for He$\alpha_z$/$w$ assuming gas motions are isotropic (left), purely radial (center) and purely tangential (right). Notation is equivalent to Fig. \ref{fig:rssims}. Purple, blue, green, yellow, and red curves show $V_\sigma=0$, $\sim70$, $140$, $250$, and $500$ km/s.}
    \vspace{0.2cm}
    \label{fig:rsaniso}
\end{figure}

Fig. \ref{fig:rsaniso} shows the results of these simulations.
The $z/w$ line ratio profiles for different gas velocities are shown in color, with equivalent notation to Fig. \ref{fig:rssims}.
In the isotropic case we find a velocity of $\sim155_{-70}^{+130}$ km/s, consistent with the measurement in Section \ref{sec:rsiso}\footnote{The small difference is due to the difference in sound speed when using a constant velocity amplitude rather than Mach number.}.
Radial gas motions have a stronger moderating effect on RS than isotropic in the inner $\sim 10$ kpc, which is expected as the line-of-sight velocity component in the cluster center is essentially radial.
Outside this radius, the effect is slightly weaker.
With purely radial gas motions, we also find a velocity of $\sim160_{-70}^{+130}$ km/s, essentially equivalent to the isotropic case.

On the other hand, pure tangential motions are less efficient at moderating RS in the cluster center (within $\sim 25$ kpc) and slightly more efficient at outer radii.
For this case, we find a lower limit on the velocity amplitude of $\gtrsim 175$ km/s.
As this limit is comparable to the broadening-measured velocity of $175\pm11$ km/s, our results slightly disfavor a scenario where small-scale motions in the inner $\sim 20$ kpc region are purely tangential at the $\sim1\sigma$ level.
Improvements to these constraints are possible with further observations and a binning scheme that targets the inner $\sim 10$ kpc of the cluster, where the differences between the three scenarios are most pronounced.
{This is achievable with further observations of the Perseus core that will improve the statistics in this inner region.} 
Future X-ray missions with even better spatial resolution \citep[such as \athena{},][]{cruise2024} will further improve the power of RS as a velocity probe in the ICM.

\subsection{Systematic uncertainties}

\textbf{Spherical symmetry:}
Our RS simulations model a spherically symmetric cluster.
However, Section \ref{sec:results} presents deviations from this assumption in Perseus.
To estimate the significance of these deviations, we compare deprojected radial profiles of the gas density, temperature, and metallicity for two wedges in Perseus covering the NW and S arms to the spherically symmetric model.
We find very little variation in the density and metallicity profiles. 
While there is some temperature variation in the inner $\sim 30$ kpc, these variations are similar to the systematic differences in temperature between \chandra{} and \xmm{}. Therefore, azimuthal variations in the cluster model are not likely to be a significant source of uncertainty in our RS simulations.
However, we cannot rule out the possibility of unresolved substructures in the ICM that could affect the RS signal.

\textbf{Atomic data:}
We investigate the systematic uncertainty due to atomic modeling by repeating our analysis with \textsc{AtomDB} version 3.0.9. 
We find no significant change in temperature, metallicity, or bulk velocity between the two versions in any region. 
However, we measure systematically higher velocity dispersions with version 3.1.3 versus version 3.0.9.
The two measurements are consistent within $1\sigma$ in all regions other than $R_1$, where the $175 \pm11$ km/s velocity dispersion measured using version 3.1.3 becomes $\sim160\pm10$ km/s, a difference of $\sim1\sigma$.
The choice of \textsc{AtomDB} version does not have a significant effect on the measured line ratios. 
More significant differences in atomic modeling may be found if we compared our \texttt{xspec} results to those from \texttt{SPEX} \citep{spex,hitomi-rs}; however, this analysis is beyond the scope of this work.

\textbf{Resonant line shape:}
Resonant scattering redistributes emission from the center of the $w$ line to its wings, making the line shape non-Gaussian \citep[e.g.,][]{churazov2010}.
In this work, we elect to model the $w$ line in the central two regions with a Voigt profile; however, we stress that this is an approximation of the actual line profile.
We also fit the $w$ line using a Gaussian component with free width in $R_1$ and $R_2$ in place of the $w$ line.
We find that the Voigt line improves the fit over the Gaussian, with an improvement in the BIC of $\sim 12$, where $>10$ is considered significant.

Moreover, modeling the $w$ line with a Gaussian results in a best-fit $z/w$ ratio of $\sim0.49\pm0.03$ in $R_1$, higher than the value of $0.45\pm0.03$ found with the Voigt profile.
This is because a Gaussian profile underestimates the flux in the $w$ line, as it does not properly model the heavier wings created by the RS effect.
While these two measurements are consistent within $1\sigma$, the derived velocity constraints are sensitive to the choice of $w$ line model.
Using a Gaussian $w$ line, we find an upper limit on the the small-scale velocity in $R_1$ of $\lesssim115$ km/s.
The lower limit on velocity in $R_2$ becomes $\sim 80$ km/s.
This highlights the importance of considering non-Gaussianity of resonant lines in regions of clusters with high optical depth.

\textbf{Multi-temperature structure:}
The $z/w$ and $\beta/w$ line ratios may be affected by the presence of multiple temperature/velocity components in the ICM.
The multi-temperature structure of Perseus is explored in detail by \citet{meunier2026}, which found a $\sim 2$ keV component in the central $\sim 60$ kpc region in addition to the $\sim4$ keV hot ICM \citep[also see][]{hitomi-temp}.
However, they found difficulty in determining whether this location is in the inner 20 kpc or the 20-60 kpc region (our $R_1$ and $R_2$) due to SSM.
I.e., the cooler component could only be identified in one of these regions at a time.
To investigate the effect of this cooler component on the RS effect, we run two different fits: one with a second in $R_1$ and another with a second component in $R_2$.
The temperature and velocity of the two regions are free, while the redshift and metallicities are tied.
We tie the components between the two arms and tie all abundances to Fe.

In $R_1$, we find a secondary component of $kT_2=1.6_{-0.1}^{+0.3}$ keV and $V_\mathrm{disp,2}=360_{-70}^{+100}$ km/s. 
In $R_2$, we find $kT_2=2.0_{-0.2}^{+0.1}$ keV and $V_\mathrm{disp,2}=410_{-70}^{+80}$. 
Consistent with the results of \citet{meunier2026}, the secondary component in $R_2$ is statistically favored $\Delta \mathrm{BIC}\approx83$.
The addition of a second component in $R_1$ has minimal effect on the $z/w$ line ratio in either $R_1$ or $R_2$.
However, the addition of a second component in $R_2$ changes the $z/w$ line ratio in $R_1$ from $\sim0.45$ to $0.46_{-0.02}^{+0.03}$ and in $R_2$ from $\sim0.36$ to $0.34_{-0.02}^{+0.01}$.
These small adjustments to the line ratio may have an effect on our RS-derived velocity constraints, particularly in $R_2$.
However, we cannot directly compare the {results of the} multi-temperature spectral model to the Monte Carlo simulations, as the latter consider only a single ICM temperature.
An improved understanding of the spatial distribution of this cool component is required in order incorporate multi-temperature models into our RS constraints.

\textbf{Charge exchange:}
In this work we do not consider the presence of charge exchange, which may have an effect on the measured line ratios \citep[see][for review]{gu2023}.
Charge exchange has yet to be conclusively detected in the ICM; however, it is likely to occur in the central regions of Perseus where cold H$\alpha$ filaments reside \citep[e.g.][]{gendron-marsolais2018}.
{Indeed, hints of charge exchange (CX) were seen in Perseus by \hitomi{} \citep{hitomi-atomic,hitomi-rs}, which mostly observed a region similar to C0. 
About $3\%$ of the total Fe He$\iota$ emission near 8.6 keV was credited to CX in these observations.
Preliminary modeling of the Fe He$\alpha$ lines using \xrism{} observations suggests a similar level of CX contamination, but with significant variations across the regions (Shefler et al. in prep.).}

\section{Conclusions}
\label{sec:conclusion}
We have presented a detailed mapping of the Perseus cluster core with high-resolution X-ray spectroscopy, analyzing a total of $\sim 830$ ks of \xrism{} observations of the Perseus cluster.
These observations cover two continuous arms extending $\sim260$ kpc NW and S of the cluster center. 
In this work, we focus on measuring the effects of resonant scattering in the ICM.

\begin{itemize}
    \item 
    We demonstrate that RS suppresses the flux of the Fe He$\alpha$ $w$ emission line in the cluster center and, for the first time, enhances the $w$ line flux in the outer regions ($\gtrsim100$ kpc from the center).
    This constitutes the first confirmation of the expected radial behavior of resonant scattering in the ICM.
    \item 
    We model the resonant $w$ line in the inner $\sim60$ kpc with a Voigt profile, finding a significant improvement in the fit ($\Delta \mathrm{BIC}\approx12$) over a Gaussian $w$ line. 
    Moreover, we find that a Gaussian profile results in significant residuals in the wings of the $w$ line.
    This further confirms the presence of RS and demonstrates the importance of considering the non-Gaussianity of resonant lines with large optical depths in the ICM.
    \item 
    We quantify the amplitude of the RS effect by measuring the emission line ratios of the optically-thin He$\alpha_z$ and He$\beta$ lines to the optically thick $w$ line.
    We find enhanced $z/w$ and $\beta/w$ ratios in the inner $\sim 60$ kpc compared to expectations for an optically thin plasma.
    Outside this radius, the line ratios are consistent with/lower than expected, further confirming the radial behavior of RS.
    \item 
    By performing Monte Carlo radiative transfer simulations tailored to the Perseus cluster, we constrain the small scale turbulent velocity to be $\sim 130_{-55}^{+115}$ km/s in the inner 20 kpc and $\gtrsim 90$ km/s in the 20-60 kpc region.
    Both constraints are consistent (within $1\sigma$ uncertainties) with the velocities measured via Doppler broadening, confirming that the measured velocity dispersions are primarily due to random, small-scale motions, rather than solely bulk flows.
    \item 
    We run a second set of RS simulations to investigate the anisotropy of small-scales gas motions in Perseus.
    We find that assuming either purely radial or isotropic motions produce consistent RS effects in the center of Perseus, within the uncertainties.
    Therefore, longer \xrism{} observations are needed to distinguish between these scenarios.
    In particular, better statistics are needed to spatially resolve the central $\sim10$ kpc region.
    Assuming purely tangential motions, the observed line ratios are reproduced by the Monte Carlo simulations if the velocity is $\gtrsim175$ km/s, which is consistent with/greater than the broadening-measured velocity dispersion.
    Therefore, our analysis slightly disfavors a scenario wherein gas motions in the inner $\sim 20$ kpc region of Perseus are purely tangential.
    However, we cannot rule out this scenario given the nontrivial difference in scales probed by Doppler broadening and resonant scattering.
\end{itemize}

In addition, we measure continuous profiles of the temperature, metallicity, bulk velocity and velocity dispersions along both arms.
We analyze the average radial profile as well as azimuthal variations between the NW and S.
\begin{itemize}
    \item 
    The gas in the S arm is systematically redshifted with respect to the NW arm, implying coherent or rotational motions between the two regions of the cluster.
    This is the product of merger-driven gas sloshing, and is consistent with the model presented in \citet{zhang2026}.
    Additionally, the velocity dispersion in $R_\mathrm{5,S}$ is higher than in $R_\mathrm{5,NW}$, likely also the product of large-scale sloshing.
    \item 
    While the plasma temperature monotonically increases to the NW, we find an isothermal region between $\sim 60$ and 260 kpc to the S, consistent with \chandra{} and \xmm{} measurements.
    This isothermal region is likely the product of gas sloshing, as the region is approximately bounded by two cold fronts.
    \item 
    The gas in $R_3$ (60-100 kpc) has a higher metallicity and velocity dispersion in the S arm than the NW.
    Both enhancements are detected at the $\gtrsim2\sigma$ level.
    This may indicate the AGN's sphere of influence is larger in the S than the NW, where the inner cold front could be limiting the effects of feedback.
    It may also be related to the bay structure in $R_\mathrm{4,S}$, which has been posited to be either a fossil AGN bubble or a Kelvin-Helmholtz instability.
    Either scenario could produce the observed enhancements compared to the NW.

\end{itemize}

Our work represents the first conclusive detection of the full effects of resonant scattering\footnote{Not including polarization effects.}: resonant line flux suppression in the dense central regions of the cluster, flux enhancement outside the center, and RS-induced non-Gaussianity of the resonant line.
We emphasize that even in \xrism{} era of high-resolution X-ray spectroscopy, multiple velocity probes are needed to understand the scale and nature of observed gas motions.
Resonant scattering is an invaluable tool to selectively measure small-scale random motions where diverging bulk flows may also be present, such as the central cool cores.
In the future, analyzing RS effects in more clusters hosting AGN could further our understandings of feedback mechanisms, e.g., how much of their energy is deposited as random motions rather than coherent ones.

\section*{Acknowledgments}

This work was supported by NASA grants 80NSSC25K0143 and 80NSSC18K1684. 
IZ acknowledges partial support from the Alfred P. Sloan Foundation through the Sloan Research Fellowship. 
Support for JZ was provided by the {\it Chandra} X-ray Observatory Center, which is operated by the Smithsonian Astrophysical Observatory for and on behalf of NASA under contract NAS8-03060.
This material is based upon work supported by NASA under award number 80GSFC24M0006.
CZ acknowledges the support of the Czech Science Foundation (GACR) Junior Star grant no. GM24-10599M.
SU acknowledges the support by JSPS KAKENHI grant numbers JP25K23398 and JP26K00741. 
SU also acknowledges support by Program for Forming Japan’s Peak Research Universities (J-PEAKS) Grant Number JPJS00420230006.

\bibliographystyle{apsrev4-1}

\bibliography{arxiv-vers}

\begin{appendix}
\setcounter{table}{0}
\renewcommand{\thetable}{A\arabic{table}}
\setcounter{figure}{0}
\renewcommand{\thefigure}{A\arabic{figure}}

\begin{figure}
    \centering
    \includegraphics[width=\linewidth]{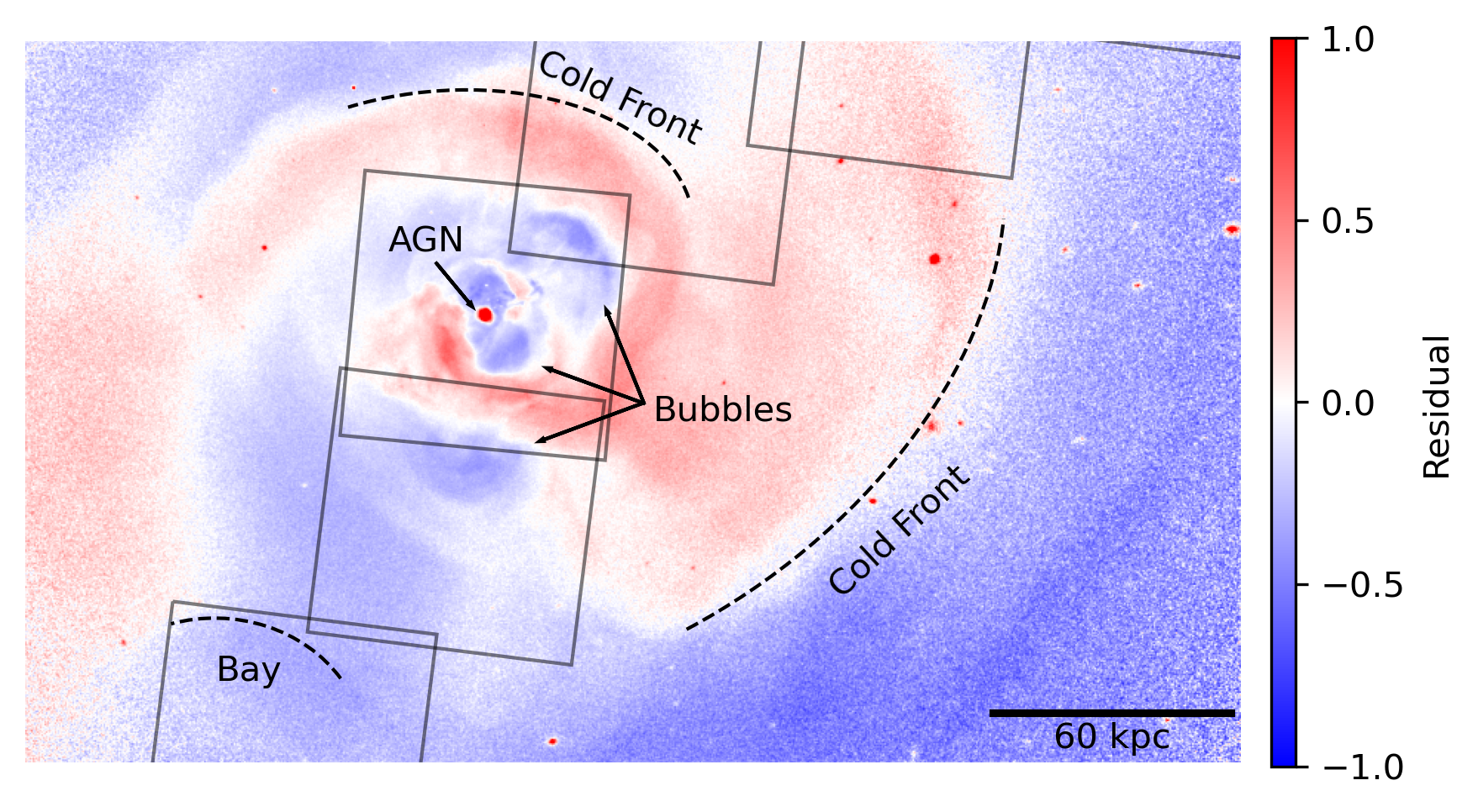}
    \caption{\chandra{} residual image ($\mathrm{Image}/\mathrm{Model}-1$) of the Perseus cluster core, with the central AGN, feedback-inflated bubbles, prominent cold fronts, and the ``bay'' structure labeled. \resolve{} FoVs are shown as gray boxes.}
    \label{fig:resid}
\end{figure}

\begin{figure}
    \centering
    \includegraphics[width=\linewidth]{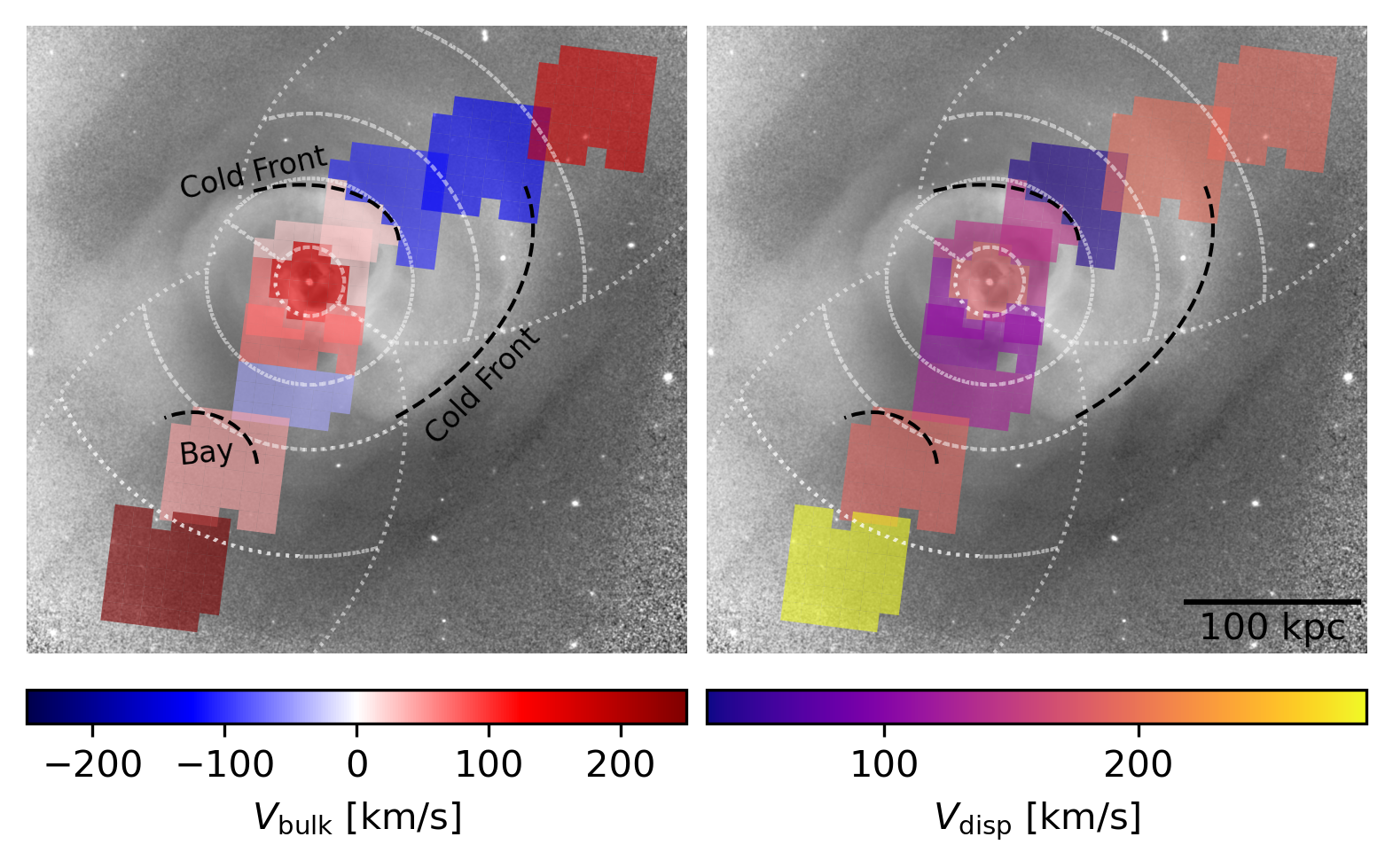}
    \caption{{Maps of the measured bulk velocity (left) and velocity dispersion (right) (when the NW and S arms are independent) overlaid on \chandra{} residual image (Fig. \ref{fig:resid}). We emphasize that the photons in each spectrum come from a region around the pixels (shown in color), with $\gtrsim95\%$ of photons originating from the on-sky regions (outlined with dotted white lines). The cold fronts and bay structure are labeled using dashed black lines.}}
    \label{fig:velmap}
\end{figure}

\section{Deprojected cluster model}
\label{app:model}

\begin{figure}
    \centering
    \includegraphics[width=\linewidth]{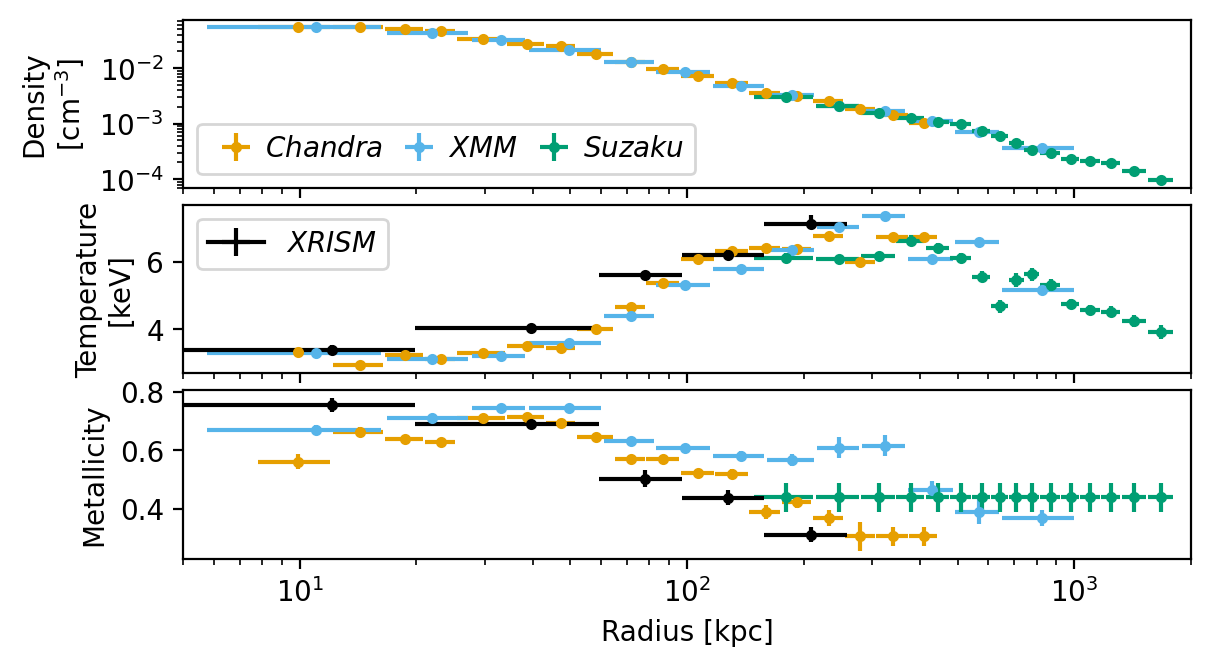}
    \caption{Deprojected profiles of electron density, gas temperature, and metallicity derived from \chandra{}, \xmm{}, and \suzaku{} data (orange, blue, and green crosses). We additionally plot our \xrism{}-measured (projected) temperature and metallicity in black.}
    \label{fig:perseusmod}
    \vspace{0.5cm}
\end{figure}

In our RS simulations, we assumed a spherically symmetric model of the Perseus cluster based on combined deprojected radial profiles derived from \chandra{}, \xmm{}, and \suzaku{} observations (Fig. \ref{fig:perseusmod}). 
We performed standard processing of the \chandra{} (obsIDs 3209, 4289, 4946--4953, 6139, 6145, 6146, 11713--11716, 12025, 12033, 12036, 12037) and \xmm{} (obsIDs 0085110101, 0085590201, 0151560101, 0204720101, 0204720201, 0305690101, 0305690301, 0305690401, 0305720101, 0305720301, 0305780101, 0405410101, 0405410201, 0673020201, 0673020301, 0673020401) data, applying the latest calibrations and standard data-reduction techniques \citep{Vikhlinin2005,churazov2003}. 
We then deprojected the spectra following \citet{churazov2003} and fitted them in XSPEC using an absorbed, single-temperature \texttt{apec} model.
We performed the spectral fitting both with the Fe abundance fixed at a constant value relative to Solar and with the Fe abundance left as a free parameter, to verify that the resulting deprojected density and temperature profiles are robust to this assumption.
The resulting radial profiles, shown in Fig. \ref{fig:perseusmod}, were then combined with the \suzaku{} measurements from \citet{urban2014}.
For the Fe abundance, we averaged the numbers from \citet{werner2013}, \citet{matsushita2013}, and \citet{urban2014}, following \citet{hitomi-rs}.

\section{AGN model constraints}
\label{app:agn}

\begin{table}[]
    \centering
    \begin{tabular}{l|llllll}
     & ObsID & (RA,Dec) [deg] & Exposure [ks] & \makecell{Heliocentric \\ Velocity [km/s]} & Region & Pixels \\
\hline \\
C0 & 000154000 & (49.9507, 41.5117) & 49 & $-24.6$ & $R_\mathrm{1}$ & 0, 2, 7, 10, 15, 17-18, 20, 25, 28, 33, 35 \\
 & & & & & $R_\mathrm{2,NW}$ & 9, 11, 13-14, 19, 21-24, 26, 34 \\
 & & & & & $R_\mathrm{2,S}$ & 1, 3-6, 8, 16, 29-32 \\
 & 000155000 & (49.9508, 41.5116) & 53 & $-24.9$ & $R_\mathrm{1}$ & 0, 2, 7, 10, 15, 17-18, 20, 25, 28, 33, 35 \\
 & & & & & $R_\mathrm{2,NW}$ & 9, 11, 13-14, 19, 21-24, 26, 34 \\
 & & & & & $R_\mathrm{2,S}$ & 1, 3-6, 8, 16, 29-32 \\
 & 101011010 & (49.9501, 41.5122) & 40 & $+24.7$ & $R_\mathrm{1}$ & 0, 2, 7, 10, 15, 17-18, 20, 25, 28, 33, 35 \\
 & & & & & $R_\mathrm{2,NW}$ & 1, 3-6, 8, 16, 29-32 \\
 & & & & & $R_\mathrm{2,S}$ & 9, 11, 13-14, 19, 21-24, 26, 34 \\
 & 101012010 & (49.9501, 41.5121) & 44 & $+24.8$ & $R_\mathrm{1}$ & 0, 2, 7, 10, 15, 17-18, 20, 25, 28, 33, 35 \\
 & & & & & $R_\mathrm{2,NW}$ & 1, 3-6, 8, 16, 29-32 \\
 & & & & & $R_\mathrm{2,S}$ & 9, 11, 13-14, 19, 21-24, 26, 34 \\
 & 102007010 & (49.9514, 41.5116) & 44 & $-25.2$ & $R_\mathrm{1}$ & 0, 2, 7, 10, 15, 17-18, 20, 25, 28, 33, 35 \\
 & & & & & $R_\mathrm{2,NW}$ & 9, 11, 13-14, 19, 21-24, 26, 34 \\
 & & & & & $R_\mathrm{2,S}$ & 1, 3-6, 8, 16, 29-32 \\
 & 102008010 & (49.9511, 41.5114) & 47 & $-25.4$ & $R_\mathrm{1}$ & 0, 2, 7, 10, 15, 17-18, 20, 25, 28, 33, 35 \\
 & & & & & $R_\mathrm{2,NW}$ & 9, 11, 13-14, 19, 21-24, 26, 34 \\
 & & & & & $R_\mathrm{2,S}$ & 1, 3-6, 8, 16, 29-32 \\
C1 & 000156000 & (49.9074, 41.5454) & 58 & $-25.2$ & $R_\mathrm{2,NW}$ & 0-8, 16, 28 \\
 & & & & & $R_\mathrm{3,NW}$ & 9-11, 13-15, 17-26, 29-35 \\
C3 & 201078010 & (49.9580, 41.4739) & 54 & $+24.0$ & $R_\mathrm{2,S}$ & 0-6, 8, 28-35 \\
 & & & & & $R_\mathrm{3,S}$ & 9-11, 13-26 \\
M1 & 000157000 & (49.8475, 41.5654) & 94 & $-25.5$ & $R_\mathrm{4,NW}$ & 0-11, 13-26, 28-35 \\
M3 & 201079010 & (50.0000, 41.4300) & 60 & $-27.8$ & $R_\mathrm{4,S}$ & 0-11, 13-26, 28-35 \\
 & 201079020 & (49.9997, 41.4299) & 62 & $-27.8$ & $R_\mathrm{4,S}$ & 0-11, 13-26, 28-35 \\
O1 & 000158000 & (49.7855, 41.5875) & 132 & $-25.9$ & $R_\mathrm{5,NW}$ & 0-11, 13-26, 28-35 \\
O3 & 201080010 & (50.0341, 41.3861) & 92 & $+24.3$ & $R_\mathrm{5,S}$ & 0-11, 13-26, 28-35 \\ 
    \end{tabular}
    \caption{Observation parameters in this analysis, including ObsID, pointing (RA, Dec), net exposure time, heliocentric velocity, and the detector regions from which spectra were extracted. We analyzed 13 observations with a total exposure time of $\sim830$ ks, from which 27 spectra were extracted.}
    \label{tab:obsidinfo}
    \vspace{0.5cm}
\end{table}

As the AGN and $R_1$ ICM normalizations are degenerate parameters, we must use \chandra{} observations to place indirect constraints on the AGN flux.
Taking advantage of \chandra{}'s spatial resolution, we remove the central AGN from the data and extract spectra from on-sky regions $R_1$ and $R_2$ using \chandra{} ObsIDs 3209, 4952, and 11714.
The spectra from each region are fit between 3.0 and 7.5 keV, modeled as an absorbed \texttt{vapec} model, where nH is fixed (see Section \ref{sec:modeling}) and all abundances are tied to Fe.
From each region we measure the flux between 4 and 6 keV (an energy range without significant emission lines) and calculate the continuum flux ratio between $R_1$ and $R_2$, $0.271\pm0.001$.
We also check other energy ranges (0.5-7.5 and 4.0-7.5 keV) and find no significant variation in the measured flux ratio.

\begin{figure}
    \centering
    \includegraphics[width=\linewidth]{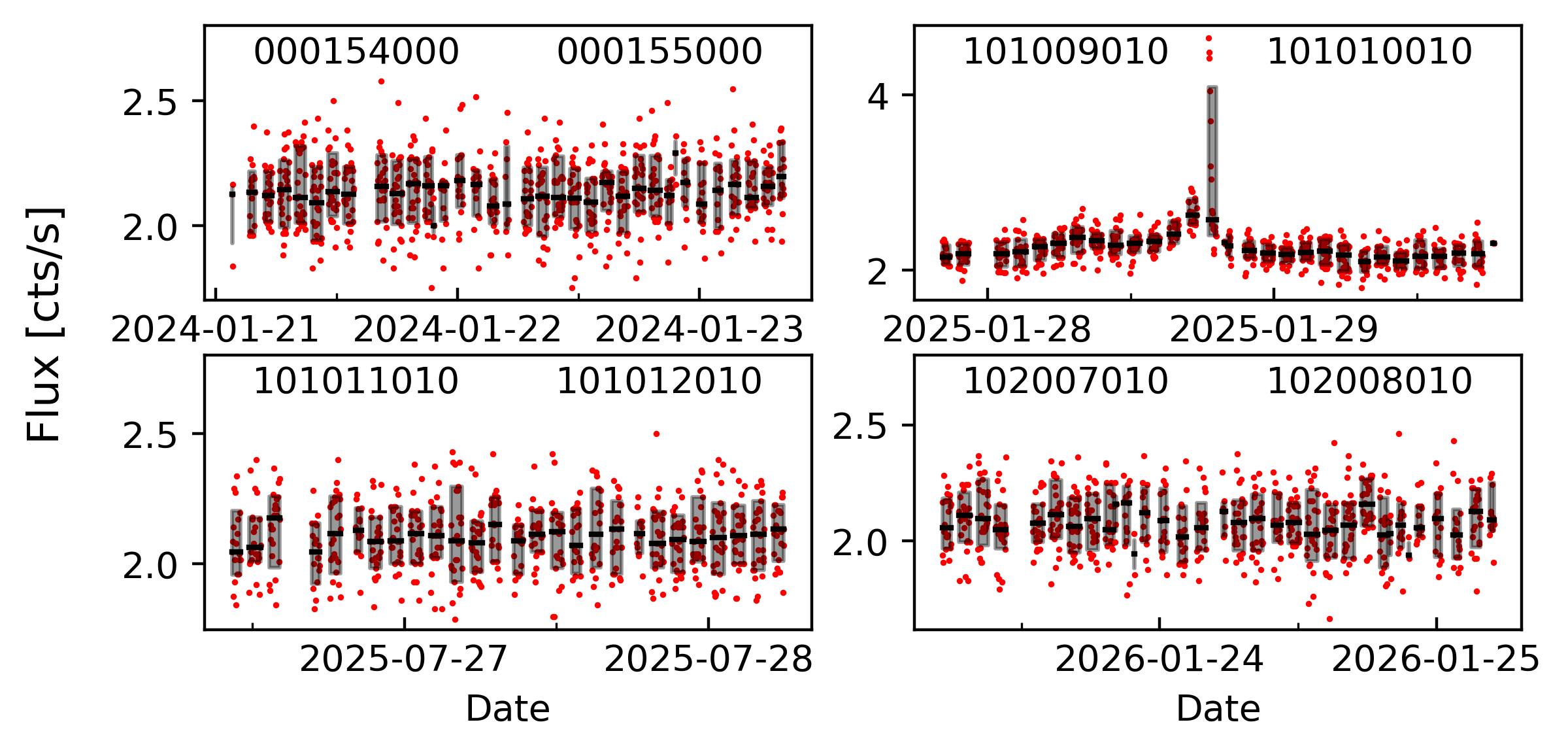}
    \caption{Full-array lightcurves for each of the central \xrism{} observations. 128-second bins are shown in red, while black shaded regions show average fluxes and $1\sigma$ standard deviations within each observation interval. Due to the presence of a significant flare, ObsIDs 101009010 and 101010010 (top right) were not included in our analysis.}
    \label{fig:C0lightcurves}
    \vspace{0.7cm}
\end{figure}

We then use this measurement of the $R_1$:$R_2$ flux ratio to constrain the AGN models.
The AGN is modeled as an absorbed powerlaw (\texttt{tbabs}$\cdot$\texttt{pegpwrlw}) with normalization parameterized as the flux at 6 keV.
To account for variations in AGN emission, we use an independent AGN model for each observation that covers the central region (ObsIDs 000154000, 000155000, 101011010, 101012010, 102007010, 102008010).
We do not consider variations in the AGN model within each observation, which is reasonable given the lack of significant variation in the lightcurves of each of these observations (Fig. \ref{fig:C0lightcurves}).
We do not include ObsIDs 101009010 and 101010010 in our dataset due to the presence of a strong AGN flare.

\begin{figure}
    \centering
    \includegraphics[width=\textwidth]{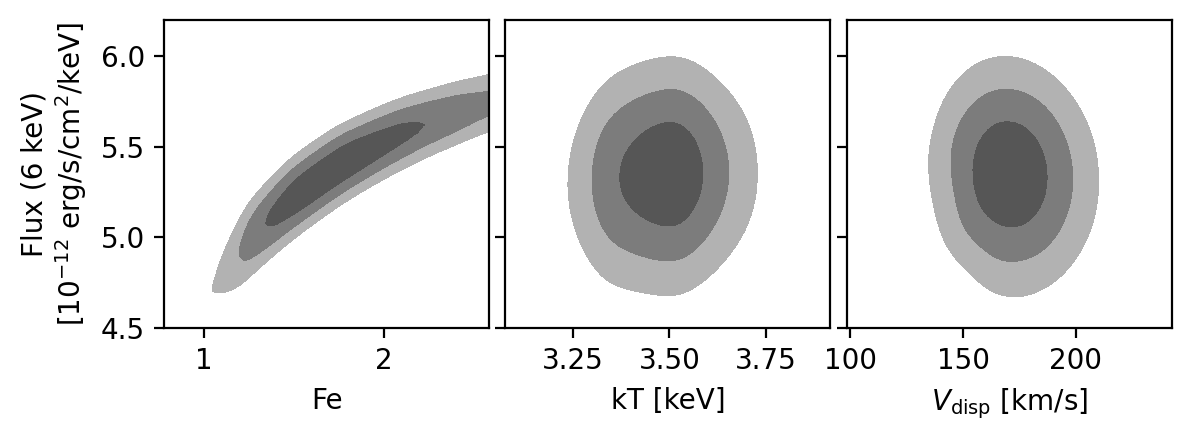}
    \caption{Constraints on the AGN flux during ObsID 000154000 compared to the Fe abundance, temperature, and velocity dispersion of $R_1$. When measuring these constraints the AGN flux is a free parameter. While the central abundance depends heavily on the AGN model, the temperature and velocity dispersion do not.}
    \label{fig:freeagn}
    \vspace{0.6cm}
\end{figure}

We initially model the data without placing any constraints on the AGN parameters.
This initial fit is biased towards a low $R_1$ normalization and high AGN flux, which produces an unphysically high central metallicity \citepalias[see also][]{xrism-pers}.
We use this unconstrained fit to measure the difference in AGN normalization between each AGN model (e.g., we find the AGN to be $\sim 9\times10^{-14}$ erg/s/cm$^2$/keV brighter in ObsID 000155000 than 000154000).

We then fix these flux differences and perform a series of fits with different fixed AGN fluxes.
This allows us to find the combination of AGN fluxes that produces the $R_1$:$R_2$ flux ratio that matches the value measured by \chandra{}.
\footnote{This is slightly different to the flux ratio method described in \citetalias{xrism-pers}, which fixed the $R_1$:$R_2$ and $R_1$:$R_3$ flux ratios before fitting. Our version consistently fixes the AGN flux throughout, ensuring the $R_1$:$R_2$ ratio is consistent with Chandra’s value. Additionally, we model the entire dataset while constraining the AGN, while \citetalias{xrism-pers} only models the inner three regions to find the AGN flux. This, in addition to the new data, may be the cause of the difference in AGN flux we derive. The 2-10 keV integrated flux of our AGN model for ObsIDs 000154000 and 000155000 is $\sim 37\times10^{-12}$ erg/s/cm$^2$, compared to the value of $\sim31\times10^{-12}$ erg/s/cm$^2$ adopted by \citetalias{xrism-pers}.}
These fluxes and the measured spectral indices (with statistical uncertainties) are listed in Table \ref{tab:agnpar}.
In our main analysis, we fix these AGN fluxes, while $\Gamma$ remains a free parameter.

To check these results, we also use the equivalent width method employed in \citetalias{xrism-pers}.
We measure the equivalent width $W_\mathrm{eq}$ of the Fe lines in the \chandra{} spectrum of $R_1$ by calculating the ratio of the Fe emission to the continuum emission between 6 and 7 keV.
Unlike the continuum flux ratio, we find that the equivalent width is heavily dependent on the energy range over which the \chandra{} spectrum is modeled.
For a broad-band fit (0.5-7.5 keV), we find $W_\mathrm{eq}\approx1.120\pm0.004$, while a 4.0-7.5 keV fit yields $W_\mathrm{eq}\approx0.823\pm0.004$.
By comparing these equivalent widths to those measured by \xrism{}, we find the AGN flux during the 000154000 ObsID can vary from $\sim 3.61$ to $4.47\times10^{-12}$ erg/s/cm$^2$/keV.
We perform another series of fits to cover this systematic uncertainty in the AGN flux.
Using the lower (upper) limit on the AGN flux, we find a central Fe abundance of $\sim 0.69\pm0.02$ ($0.91\pm0.03$), compared to our baseline central abundance of $\sim 0.76$.
No other parameters, including the measured line ratios, are significantly affected by this change in AGN flux.
To further demonstrate this, we plot the AGN flux constraints (ObsID 000154000) from our free-AGN fit against the central Fe abundance, temperature, and velocity dispersion constraints in Fig. \ref{fig:freeagn}.

\renewcommand{\arraystretch}{1.5}
\begin{table}[]
    \centering
    \begin{tabular}{l|ll }
    ObsID & $\Gamma$ & \makecell[l]{Flux (6 keV) \\ \ [$10^{-12}$ erg/s/cm$^{2}$/keV]} \\
    \hline
    000154000 &
    $1.82 \pm 0.04$ &
    $3.99$ \\
    000155000 &
    $1.84_{-0.08}^{+0.04}$ &
    $4.08$ \\
    101011010 &
    $1.82_{-0.09}^{+0.05}$ &
    $3.87$ \\
    101012010 &
    $1.88_{-0.08}^{+0.04}$ &
    $3.82$ \\
    102007010 &
    $1.87 \pm 0.05$ &
    $3.59$ \\
    102008010 &
    $1.98_{-0.08}^{+0.04}$ &
    $3.82$ \\
    \end{tabular}
    \caption{AGN Model parameters for each central ObsID.}
    \label{tab:agnpar}
\end{table}

\end{appendix}

\end{document}